\documentclass[12pt]{article}
\usepackage{amsmath} 
\usepackage{amsfonts} 
\usepackage{graphicx} 
\usepackage{microtype} 
\usepackage[numbers,square,sort&compress]{natbib} %
\usepackage[parfill]{parskip}
\def\Eq#1{Eq.~\ref{#1}}
\def\Fig#1{Fig.~\ref{#1}}
\title{Gravitatational Waves in G4v\footnote{\copyright  2015 Carver Mead}} 
\author{Carver Mead}

\begin{document}
\maketitle

\begin{abstract}
\noindent
Gravitational coupling of the propagation four-vectors of matter wave functions is formulated
in flat space-time.  Coupling at the momentum level rather than at the ``force-law'' level greatly simplifies
many calculations.  This locally Lorentz-invariant approach (G4v) treats electromagnetic
and gravitational coupling on an equal footing.  Classical mechanics emerges from the incoherent aggregation
of matter wave functions.  The theory reproduces, to first order beyond Newton, the standard GR results for
Gravity-Probe B, deflection of light by massive bodies, precession of orbits, gravitational red shift, and total
gravitational-wave energy radiated by a circular binary system.  Its predictions differ markedly from GR for
the gravitational-wave radiation patterns from rotating massive systems, and for
the LIGO antenna pattern.  G4v predictions of total radiated energy from highly eccentric Kepler systems are
slightly larger than those of similar GR treatments.
A detailed treatment of the theory is in preparation.  However the generation
and detection of gravitational waves is exactly the same as the corresponding treatment for electromagnetic waves
given in {\it Collective Electrodynamics}\cite{Mead2002} 
(hereinafter referred to simply as {\bf CE}) and therefore separable from the material in preparation.  It therefore seems advisable to make
the gravitational-wave material available, since its predictions should be testable as data from Advanced LIGO becomes
available over the next few years.  The presentation is somewhat more detailed than would be ``normal,'' simply to
make the approach clear and accessible to non-specialists.
\end{abstract}
\vfil\eject
\tableofcontents

\section{Motion Masses due to Gravitational Waves}
\label{MassMotion}
In G4v, the gravitational vector potential $\vec A$ is the direct measure of how the local frame of reference is affected 
by the movement of distant matter.  If we choose our coordinate system for local physics
to be defined by the universe at large, the moving matter will have a very small influence on
the coordinate system.  In that global coordinate system, a free-floating mass in a uniform gravitational
scalar potential will have a velocity 
\begin{equation}
\vec v=c\vec A_\perp
\label{vA}  
\end{equation}
Let us consider two masses, $m_1$ located at $\vec r_1$ and $m_2$ located at $\vec r_2$
relative to the binary source.  The relative position of 
the two masses is $\vec l=\vec r_2-\vec r_1$.
Because of the vector potential of the source masses, the positions $\vec r_1$ and $\vec r_2$
will be functions of time.  Each mass will move around a little orbit, periodic in time.
If the two masses are part of a gravitational-wave detector and the binary source is
at astronomical distance,  the spacing between the two masses 
will be vanishingly small compared to either individual distance.  
For that reason the direction and amplitude of the two motions will be the same,
only executed at different times due to the difference in arrival time $\delta t=(r_2-r_1)/c$ 
of the potentials at the two masses\footnote{Throughout this document, when a vector name
is used without the vector arrow above it, it means the modulus (``length'') of the vector,
and the vector name with a hat over it means a unit vector in its direction.}.
\begin{equation}
\begin{aligned}
\vec r_1(t)-\vec r_1(0)&=\ \vec r_2(t+\delta t)-\vec r_2(\delta t)
\label{r12}
\end{aligned}
\end{equation}

If the wavelength of the gravitational wave is much longer than the distance between the two masses
\begin{equation}
\begin{aligned}
\vec r_2(t+\delta t)
&\approx\ \vec r_2(t)+\frac{\partial \vec r}{\partial t}\delta t
=\ \vec r_2(t)+\vec v\ \delta t
\label{dr2}
\end{aligned}
\end{equation}
so the vector $\vec l$ from $m_1$ to $m_2$ is
\begin{equation}
\begin{aligned}
\vec l&=\vec r_2(t)-\vec r_1(t)\approx\ \vec l_0-\vec v\ \delta t
\label{vecl}
\end{aligned}
\end{equation}
where $\vec l_0=\vec r_2(\delta t)-\vec r_1(0)$ is not a function of time.

Using $\hat R$ as the unit vector in the direction of $\vec r_1$ or $\vec r_2$, the distance between the two masses thus becomes
\begin{equation}
\begin{aligned}
l&\approx\ l_0-\vec v\cdot\hat l\ \delta t\cr
&\approx\ l_0-c\vec A \cdot\hat l\ \delta t\cr
&\approx\ l_0-c\vec A \cdot\hat l\ \frac{\hat R\cdot\vec l}{c}\cr
&\approx\ l_0-\left(\vec A \cdot{\hat l}\right)\left(\hat R\cdot\vec l\right)
\label{l}
\end{aligned}
\end{equation}
The fractional change in length between the two masses thus becomes
\begin{equation}
\begin{aligned}
\frac{\delta l}{l}&\approx
-\left(\vec A\cdot{\hat l}\right)\left(\hat R\cdot\hat l\right)
\label{dl}
\end{aligned}
\end{equation}
We can understand this formula intuitively in the following way:\\
The first term arises because the velocity of each individual mass is in the direction of $\vec A$.
The second term expresses the fact that the position of each mass is the velocity times the time.
Thus the difference in the positions will be the difference in vector potential times the difference in time of arrival.

\section{Gravitational Wave Radiation}
In G4v, propagating waves are characterized by the Green's function of the source energy-momentum 4-vector,
which takes the place of the charge-current 4-vector of electrodynamics.  In other respects the calculations are exactly
the same as their electrodynamic counterparts, which have been set forth in {\bf CE}. 
In the far-field, which is all we will consider in what follows, the transverse vector potential is the only contributor to
the wave's interaction with other matter, leading to substantial simplification of the calculations, as detailed in Section \ref{AppendixTT}.
\begin{equation}
\begin{aligned}
{\vec A}=\frac{G}{c^3}\sum_i\frac{\vec p_i}{r_i}
\approx\frac{G}{c^3}\sum_i\frac{m_i \vec v_i}{r_i}
\end{aligned}
\label{Green4}
\end{equation}

\subsection{Circular Binary Source}
\begin{figure}[!ht]
\begin{center}
\includegraphics[height=10cm]{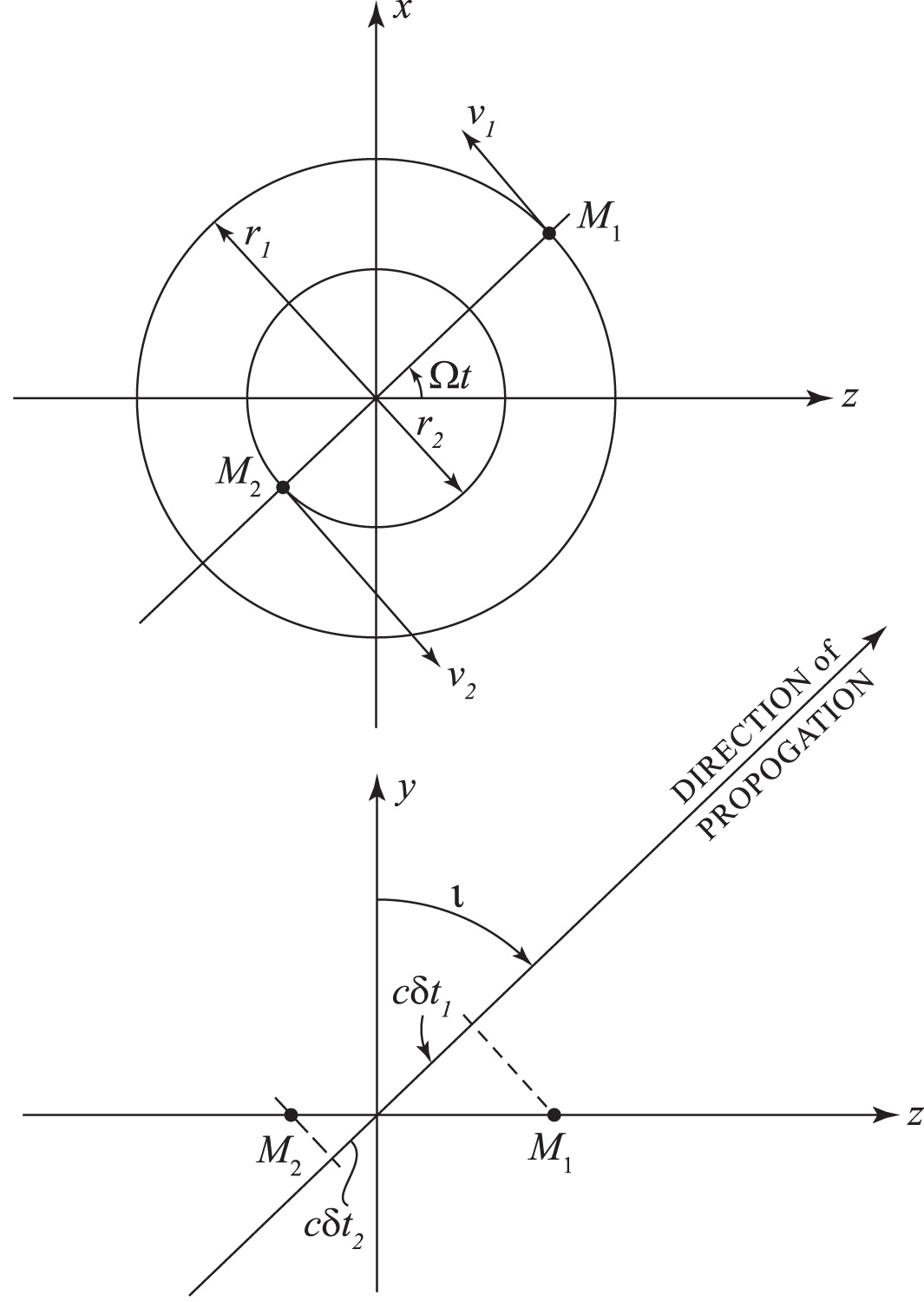} 
\caption{Top and side views of binary gravitational wave source. 
Masses $M_1$ and $M_2$ are orbiting the $y$ axis (axis of rotation). 
The velocities are $v_1$ and $v_2$, and the radii of the orbits are $r_1$ and $r_2$. 
Observation is done at a point located at a distance $R\gg r_1$, in the $y-z$ plane
at an angle $\iota$ from the $y$ axis.  The angle $\iota$ is called the {\bf inclination}
of the binary with respect to the line of sight.  The time difference $\delta t_1-\delta t_2$ is
calculated from the distances $c\, \delta t_1$ and $c\,\delta t_2$ along the path to the observer.
\label{Binary}}
\end{center}
\end{figure}

We first consider the simplest possible source of gravitational waves---a binary system 
consisting of two masses $M_1$ and $M_2$ orbiting their common
center of mass with angular frequency $\Omega$, as shown in \Fig{Binary}. 
Although simple, this model is an excellent approximation of many real binary systems.
Mass $M_1$ has velocity $v_1$ at orbital radius $r_1$
and Mass $M_2$ has velocity $v_2$ at orbital radius $r_2$, where $M_1 r_1 = M_2 r_2$.

The $z$ positions of the two masses are
\begin{equation}\begin{aligned}
z_{1} &=r_1 \cos{\Omega t}\cr
z_{2} &=-r_2 \cos{\Omega t}
\end{aligned}
\label{zpos}
\end{equation}
Thus the time differences $\delta t_1$ and $\delta t_2$ along the path to the observer between the Green's functions 
of the two masses is\footnote{Because observation is done at a distant point in the $y-z$ plane,
the difference in x makes a negligible contribution to the delay.}
\begin{equation}
\begin{aligned}
c\, \delta t_1 &=-z_1 \sin{\iota}=-r_1 \cos{\Omega t} \sin{\iota}\cr
c\, \delta t_2 &=-z_2 \sin{\iota}=r_2 \cos{\Omega t} \sin{\iota}
\end{aligned}
\label{dt}
\end{equation}
where the reference position ($\delta t=0$) is taken at the center of the binary.  We restrict this simple example to velocities $v\ll c$.
The z-directed velocities of $M_1$ and $M_2$ are

\begin{equation}
\begin{aligned}
v_{z1} &=\frac{\partial z_1}{\partial t}
=-\Omega r_1\sin{\Omega t}\cr
v_{z2} &=\frac{\partial z_2}{\partial t}
=\Omega r_2\sin{\Omega t}
\end{aligned}
\label{vz}
\end{equation}
The velocities $v_1$ and $v_2$ with respect to the observer are
those given above multiplied by $\sin{\iota}$.

\subsubsection{Vector Potential} 
The vector potential $\vec A$ at an observation point at distance $R\gg r_1$
is the sum of the Green's functions of the individual masses (\Eq{Green4}):

The $x$ components of the momenta are
\begin{equation}\begin{aligned}
p_{x1} \approx M_1v_{1x}&=M_1\Omega\, r_1 \cos{\Omega t}=M\Omega\, r\cos{\Omega t}\cr
p_{x2} \approx M_2v_{2x}&=-M_2\Omega\, r_2 \cos{\Omega t}=-M\Omega\, r\cos{\Omega t}
\end{aligned}
\label{xvel}
\end{equation}
The $z$ components of the momenta are
\begin{equation}\begin{aligned}
p_{z1} \approx M_1v_{1z}&=-M_1\Omega\, r_1 \sin{\Omega t}=-M\Omega\, r \sin{\Omega t}\cr
p_{z2} \approx M_2v_{2z}&=M_2\Omega\, r_2 \sin{\Omega t}=M\Omega\, r \sin{\Omega t}
\end{aligned}
\label{zvel}
\end{equation}
and the $z$ positions are given by \Eq{zpos}.  \\
The last form is possible because $M_1 r_1=M_2 r_2=Mr$ and $p_1=-p_2$.\\
We will often use the quantities $r=(r_1+r_2)/2$ and $M=2M_1M_2/(M_1+M_2)$,\\
which reduce to the obvious mass and radius of the symmetric system.\\
For historic reasons, most treatments use the {\bf reduced mass} $\mu=M_1M_2/(M_1+M_2)=M/2$.

We calculate the the $x$ and $z$ components of $\vec A$ separately:
\subsubsection{${\bf A}_x$ component}
\noindent
From  \Eq{Green4}
\begin{equation}
A_x(R,t)\approx \frac{G}{c^3}\left(
\ \frac{p_{x1}\left( t-\delta t_1\right)}{R}
\ +\ \frac{p_{x2}\left( t-\delta t_2\right)}{R}
\right)
\label{Artd}
\end{equation}
where we have neglected the contribution of the $x$ location to the separation
since $R\gg r_1$, and have reset the zero of time to the arrival time at $R$.  Using \Eq{xvel}, 
\begin{equation}
\begin{aligned}
A_x(R,t)\approx {\frac{GM\Omega\, r}{R\, c^3}} \Big(
&\cos{\Omega{\left( t-\delta t_1\right)}}
\ -\ {\cos{\Omega{\left( t-\delta t_2\right)}}}\Big)\cr
= {\frac{GM\Omega\, r}{R\, c^3}} \Big(
&\cos{\Omega t}\ \cos{\Omega \delta t_1}+ \sin{\Omega t}\ \sin{\Omega \delta t_1}\cr
-&\cos{\Omega t}\ \cos{\Omega \delta t_2}- \sin{\Omega t}\ \sin{\Omega \delta t_2}\Big)\cr
= \frac{GM\Omega\, r}{R\, c^3} &\Big(\cos{\Omega t}\ (\cos{\Omega \delta t_1}-\ \cos{\Omega \delta t_2})
+\sin{\Omega t}\  (\sin{\Omega \delta t_1}-\ \sin{\Omega \delta t_2})\Big)
\label{Artd1a}
\end{aligned}
\end{equation}
All $\Omega \delta t\ll 1$, so the $\cos$ terms may be neglected to first order and the $\sin$ terms approximated by their arguments:
\begin{equation}
\begin{aligned}
A_x(R,t)&\approx\frac{GM\Omega\, r}{R\, c^3} \sin{\Omega t}\  \Big({\Omega \delta t_1}-{\Omega \delta t_2}\Big)\cr
&\approx\frac{GM\Omega^2\, r}{R\, c^3} \sin{\Omega t}\ 
\Big(-\frac{r_1}{c} \cos{\Omega t} \sin{\iota}-\frac{r_2}{c} \cos{\Omega t} \sin{\iota}\Big)\cr
&\approx-\frac{GM\Omega^2\, r}{R\, c^4} \sin{\Omega t}\cos{\Omega t}\,\sin{\iota}  \Big(r_1  + r_2  \Big)\cr
&\approx-\frac{GM\Omega^2\, r^2}{R\, c^4} \sin{2\Omega t}\,\sin{\iota}
\label{Artd2}
\end{aligned}
\end{equation}
where the second form uses $\delta t_1$ and $\delta t_2$ from \Eq{dt} and $r=(r_1+r_2)/2$.

The result can be more easily understood when the individual contributions 
are spelled out:
\begin{equation}
A_x\approx -{\frac{GM}{R\, c^4}}\cdot{\frac{\Omega\, r}{c}}
\cdot{\Omega\, r}\cdot {{\sin{2\Omega t}}}\cdot \ \sin{\iota}
\label{propsol}
\end{equation} 
The negative sign merely indicates the phase of the sinusoidal signal due to our chosen origin of the rotation.
The first term is the fraction of the frame of reference contributed by each mass $M$ at distance $R$.  
The second term is the phase shift across the orbit---if both contributions were in phase they would cancel
out at large distances and there would be no far-field radiation.  The third term is the velocity of each mass,
which it has to impart to the frame of reference.
The fourth term is the time dependence with which the masses have
momentum in the $x$ direction and the fifth term is the dependence on inclination angle $\iota$ of the binary system
relative to the direction of observation.

The $x$-directed field ${\cal E}_x$ is given by

\begin{equation}
{\cal E}_x=-\frac{\partial A_x}{\partial t}
\approx \frac{GM\Omega^3\, r^2}{R\, c^4}\cos{2\Omega t}\ \sin{\iota}
\label{propsol1}
\end{equation} 

The Transverse Theorem (Section \ref{AppendixTT}) tells us that only the vector potential transverse to the direction of propagation
contributes to interaction with matter.
In the coordinate system we are using, the components of the transverse vector potential $A_\perp$ are
$A_x$ and $A_\iota$.
\begin{equation}
\begin{aligned}
&A_\iota(R,t)=A_z \cos{\iota}\approx -\frac{GM\Omega^2\, r^2}{R\, c^4}\
\cos{2\Omega t}\ \sin{\iota} \cos{\iota}\cr
&A_\iota\approx -h_0\cos{2\Omega t}\ \sin{\iota} \cos{\iota}
\qquad{\rm where}\qquad h_0= \frac{GM\Omega^2\, r^2}{R\, c^4}
\label{Atheta}
\end{aligned}
\end{equation}
The $\iota$-directed field ${\cal E}_\iota$ is given by
\begin{equation}
\begin{aligned}
{\cal E}_\iota=-\frac{\partial A_\iota}{\partial t}
&\approx \frac{GM\Omega^3\, r^2}{R\, c^4}\sin{2\Omega t}\  \sin{2\iota}
\label{Etheta}
\end{aligned}
\end{equation} 

\subsubsection{Radiation Pattern Comparison with GR}
We are now in a position to make direct comparison of the radiation patterns predicted by 
four-vector gravity (G4v) and GR.  
\begin{figure}[!ht]
\begin{center}
\includegraphics{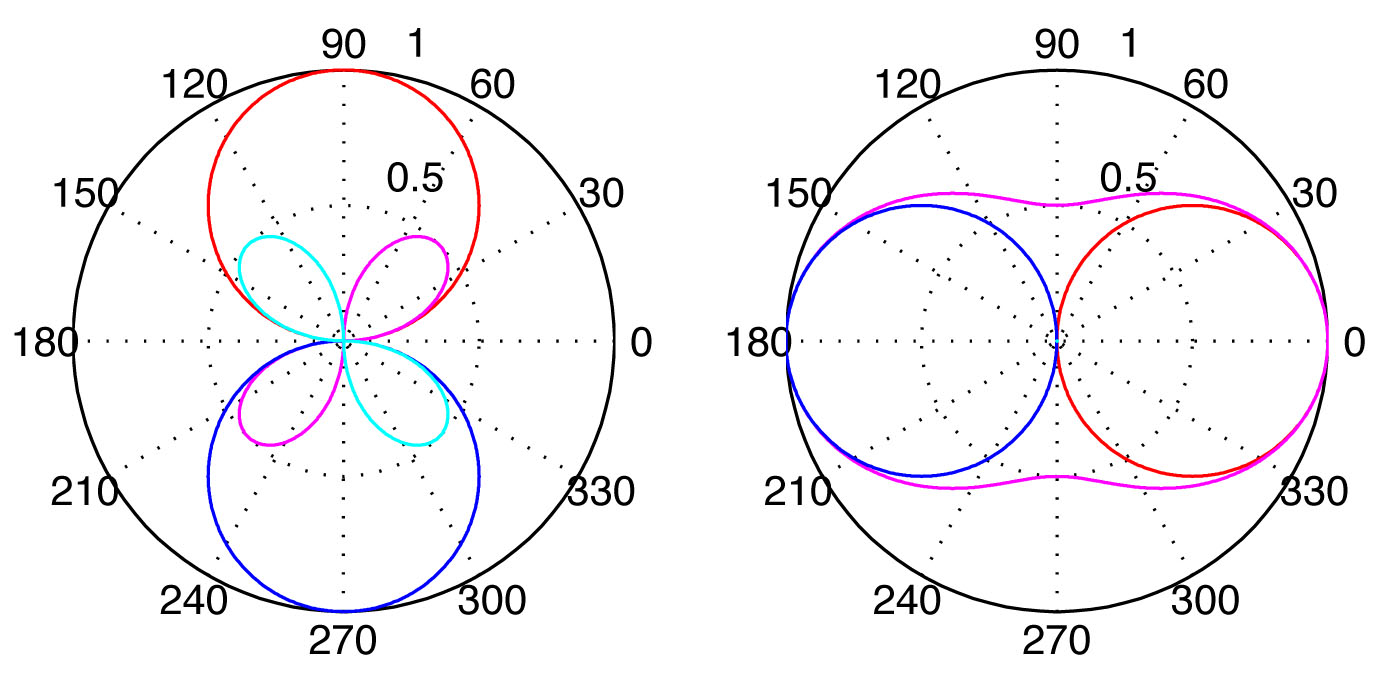} 
\caption{Comparison of quadrupole radiation patterns for G4v (left) and GR (right). 
Amplitude is plotted in the radial direction as a multiple of $h_0$ 
{\it vs} inclination  $\iota$.  Red-blue is the $A_x$ polarization for G4v, and  
the $A_+$ polarization for GR.  Magenta-cyan is the $A_\iota$ polarization for G4v, and  
the $A_\times$ polarization for GR.  
The phase of the red and magenta lobes is 180 degrees from that of the blue and cyan lobes. 
The zero of inclination is when the source rotation axis is pointing toward the observer.
\label{GRvsG4vRad}} 
\end{center}
\end{figure}

Both theories predict radiation at twice the rotation frequency of the source.
Both theories have two polarizations in phase quadrature, with the same amplitudes $h_0$: 
\begin{equation}
\begin{aligned}
GR:\qquad &h_+ = \frac{h_0}{2}\left(1 + \cos^2{\iota}\right)\, \cos{2\Omega t}{\qquad}
&h_\times = h_0 \cos{\iota}\ \sin{2\Omega t}\cr
G4v:\qquad &A_\iota\, =\, h_0\, \left(\sin{\iota}\, \cos{\iota}\right)\ \cos{2\Omega t}
&A_x=h_0 \sin{\iota}\ \sin{2\Omega t}
\label{Acompare}
\end{aligned}
\end{equation}

The amplitudes of the two radiated polarizations are, however, quite different functions of the inclination angle $\iota$.
Plots of the two polarization amplitudes as functions of inclination $\iota$ are shown 
for both GR and G4v in \Fig{GRvsG4vRad}.
The GR polarizations are ``stretch-squeeze'' tensors, while the G4v polarizations are ordinary vectors.
GR predicts non-zero amplitude at all inclinations, with a maximum on the rotation axis ($\iota=0$).  G4v 
predicts zero amplitude on the rotation axis, with maximum amplitude normal to the rotation axis ($\iota=90^\circ$). 
This is a strong prediction of G4v; one of the few cases where its predictions differ from those of GR at first order.

\subsubsection{Field Energy}
The gravitational vector potential is dimensionless.

\noindent
The gravitational field $\vec{\cal E}=-\nabla{\cal V}-\partial\vec A/\partial t$:
has the units of inverse time.  

\noindent
In the electromagnetic case, the electric field
times the quantity of charge gives the force.  In the G4v formulation of gravitation, the field
$\cal E$ times the quantity of matter $Q=mc$ gives the mass times the acceleration, which is the force.  

\noindent
Thus far the parallel between the two works well.

In the electromagnetic case, the field energy per unit volume $W_v$ of a propagating wave is
given by:

\begin{equation}
\begin{aligned}
{W_v}=\epsilon_0 \left<\vec{\cal E}^2\right>
\label{Wve}
\end{aligned}
\end{equation}
What is the gravitational equivalent of $\epsilon_0$?

\noindent
The potential $\cal V$ of a single charge $q$ charge is

\begin{equation}
\begin{aligned}
{\cal V}=\frac{q}{4\pi\epsilon_0 r}
\label{Ve}
\end{aligned}
\end{equation}

In the G4v formulation of gravitation, the potential is
\begin{equation}
\begin{aligned}
{\cal V}=\frac{G}{c^3}\frac{mc^2}{r}
=\frac{G}{c^2}\frac{mc}{r}
=\frac{G}{c^2}\frac{Q}{r}
\end{aligned}
\label{Vg}
\end{equation}
The correspondence in coupling constants is therefore
\begin{equation}
\begin{aligned}
\frac{1}{4\pi\epsilon_0}\quad\Rightarrow\quad \frac{G}{c^2}
\qquad{\rm or}\qquad 
\epsilon_0\quad\Rightarrow\quad \frac{c^2}{4\pi G}
\label{coup}
\end{aligned}
\end{equation}
Therefore the field energy per unit volume $W_v$ of a propagating wave is
just that due to the transverse field
\begin{equation}
\begin{aligned}
W_v&= \frac{c^2}{4\pi G}
\Big(\left<{\cal E}_x^2\right>+\left<{\cal E}_\iota^2\right>\Big)
\label{Wvg}
\end{aligned}
\end{equation}
The power per unit area $S$ carried by the wave is just this energy density 
multiplied by the velocity $c$
\begin{equation}
\begin{aligned}
S&=c\,W_v=\frac{c^3}{4\pi G}
\Big(\left<{\cal E}_x^2\right>+\left<{\cal E}_\iota^2\right>\Big)
\label{Pa}
\end{aligned}
\end{equation}

\subsubsection{Total Radiated Power}
As in the electric dipole case, the total power radiated $P$ is obtained by integrating
$S$ over the sphere of radius $R$
\begin{equation}
\begin{aligned}
P&=\int_{0}^{\pi}S\ 2\pi R \sin{\iota}\ Rd\iota\cr
&=\frac{ R^2 c^3}{2 G}\ \int_{0}^{\pi}
\Big(\left<{\cal E}_x^2\right>+\left<{\cal E}_\iota^2\right>\Big)\sin{\iota}\ d\iota
\label{Pa1}
\end{aligned}
\end{equation}

From\Eq{propsol1} we have
\begin{equation}
{\cal E}^2_x
\approx \frac{4G^2M^2\Omega^6\, r^4}{R^2\, c^8}\cos^2{2\Omega t}\ \sin^2{\iota}
\label{E2x}
\end{equation} 

From\Eq{Etheta} we have
\begin{equation}
\begin{aligned}
{\cal E}^2_\iota
&\approx \frac{G^2M^2\Omega^6\, r^4}{R^2\, c^8}\sin^2{2\Omega t}\ \sin^2{2\iota}\cr
\label{E2theta}
\end{aligned} 
\end{equation} 

Because $\left<\cos^2{2\Omega t}\right>=\left<\sin^2{2\Omega t}\right>=1/2$, \Eq{Pa1} 
for the average power becomes
\begin{equation}
\begin{aligned}
P&\approx{\frac{GM^2\Omega^6\, r^4}{4c^5}}\int_0^{\pi}
\Big(4\sin^2{\iota}+\sin^2{2\iota}\Big)\sin{\iota}\ d\iota 
={\frac{GM^2\Omega^6\, r^4}{c^5}}\cdot\frac{8}{5} 
\label{P2}
\end{aligned}
\end{equation}
which, using the average mass $M$, agrees with \Eq{psiintegral} when the eccentricity $\epsilon=0$.

 \subsection{Eccentric Binary Source}
As we have seen for the circular binary, the Green's function for the radiated gravitational vector potential
$\vec A$ at an observation point at distance $R\gg r$
is the sum of the Green's functions of the individual masses (\Eq{Green4}).
Of course we must use the retarded Green's functions of the two masses: 
\begin{equation}
\begin{aligned}
\vec A(R,t)&= \frac{G}{c^3}\left(\ \frac{\vec p_{1}\left( t-\delta t_1\right)}{R}\ +\ \frac{\vec p_{2}\left( t-\delta t_2\right)}{R}\right)
\approx\frac{G}{Rc^3}\Big({\vec p\left( t\right)}\ -\ {\vec p\left( t-\delta t\right)}\Big)\cr
&\approx \frac{G}{Rc^3}\left( \frac{\partial \vec p}{\partial t}\,\delta t\right)
= \frac{G}{Rc^3}\frac{\partial \vec p}{\partial t}\ \frac{\vec d\cdot\hat s}{c}
= \frac{G}{Rc^4}\frac{\partial \vec p}{\partial t}(\vec d\cdot\hat s)
\end{aligned}
\label{Artd1}
\end{equation}
where $\vec d$ is the vector distance from $m_1$ to $m_2$ and
$\delta t=\delta t_1-\delta t_2$ is the delay of the signal from $m_2$ relative to that from $m_1$
in the direction of propagation $\hat s$. 
We have neglected the contribution of the orbital location to the amplitude since $R\gg r_1$
and recognized that $\vec p=\vec p_1=-\vec p_2$.
For the final form we have made the approximation
that the time delay $\delta t$ across the orbit is small compared with the period of any radiated harmonic. 
In Peters\&Mathews\cite{Mathews63}, following the derivation of Landau\&Lifshitz\cite{LLQW}, this is one of the assumptions of
their {\bf quadrupole approximation}.

The gravitational field $\vec E$ is given by the time derivative of the vector potential:
\begin{equation}
\begin{aligned}
\vec E&= - \frac{\partial \vec A}{\partial t}
= -\frac{G}{Rc^4}\,\frac{\partial}{\partial t}\!\left(\frac{\partial \vec p}{\partial t}(\vec d\cdot\hat s)\right)&\cr
E_s&=\vec E\cdot\hat s \qquad\qquad E_\perp^2=E^2-E_s^2
\end{aligned}
\label{EvecEperp}
\end{equation}
where $E_s$ is the component of the field in the direction of propagation 
and $E_\perp$ is the component transverse to the direction of propagation.
   
\subsubsection{Total Radiated Power}
By the Transverse Theorem, the far-field vector potential is purely transverse, 
and therefore only the transverse component of the field $E_\perp$ contributes to the radiated power. 
From \Eq{Pa}, the instantaneous far-field power $P$ per unit area propagating away from the source is:
\begin{equation}
\begin{aligned}
\frac{\partial P}{\partial\, {\rm area}}&=\frac{c^3}{4\pi G}\,E_\perp^2 \qquad\Rightarrow\qquad 
\frac{\partial P}{\partial\varOmega}&=\frac{R^2 c^3}{4\pi G}\,E_\perp^2 
\label{Pa1a}
\end{aligned}
\end{equation}
where the final form
recognizes that the unit area $d\, {\rm area}=R^2 d\varOmega$, where $\varOmega$ is the solid angle in the
direction $\hat s$ from the source (not to be confused with the rotational frequency $\Omega$ in the previous section).\\
\Eq{Pa1a} is the G4v equivalent of Peters \& Mathews Eq. 3.

The {\bf line of apsides} passes through the foci of the two ellipses defining the trajectories of $m_1$ and $m_2$, 
and through the center of momentum of the system, and forms
the $\theta=0$ axis of the spherical coordinate system.  We adopt the same coordinate system used for the circular binary,
so the line of apsides is the $z$ axis.  The two ellipses lie in the same plane, which forms the $x-z$ plane of the
coordinate system.  The $y$ axis of the coordinate system is normal to the plane of the ellipses and passes through the
common focus of the two ellipses, which is the center of momentum.  The coordinates of the two masses $m_1$ and
$m_2$ derived above are given in this coordinate system with the origin of the orbital phase angle $\psi$ being the $z$ axis.  The vector $\vec r$
and the momenta of the two bodies all lie in the plane of the orbits, and therefore $\partial\vec p/\partial t$ lies in this plane as well.  

Using the same nomenclature as Peters \& Mathews,  we take $a$ as the
semi-major axis of the elliptical orbit defined by the distance between the two masses.  In the Cartesian basis the Kepler orbital
equation is:
\begin{equation}
\begin{aligned}
\vec r&=\frac{a(1-\epsilon^2)}{1+\epsilon\cos{\psi}}
\left\lbrace {\sin{\psi}},0,{\cos{\psi}}\right\rbrace\qquad\qquad \vec d=2\vec r\
\end{aligned}
\label{Krvspsi}
\end{equation}
and the angular velocity is:
\begin{equation}
\begin{aligned}
\frac{\partial  \psi}{\partial t}&=\frac{\sqrt{a(1-\epsilon^2)(m_1+m_2)G}}{r^2}
\end{aligned}
\label{Kdpsidt}
\end{equation}
We check the orbital period $P_b$ for consistency:
\begin{equation}
\begin{aligned}
P_b=\int_0^{2\pi}\frac{1}{\frac{\partial  \psi}{\partial t}}\,d\psi=2\pi\sqrt{\frac{a^3}{G(m_1+m_2)}}\cr
\end{aligned}
\label{randdpsidt}
\end{equation}
in agreement with standard treatments.  The momentum relations are thus
\begin{equation}
\begin{aligned}
\vec p&= m\frac{\partial \vec r}{\partial t}= m\frac{\partial \vec r}{\partial \psi}\frac{\partial \psi}{\partial t}\qquad\qquad
\frac{\partial \vec p}{\partial t}= \frac{\partial \vec p}{\partial \psi}\frac{\partial \psi}{\partial t}
\end{aligned}
\label{dvecpdt}
\end{equation}
Expressing the unit vector in the direction of propagation $\hat s$ in spherical coordinates 
\begin{equation}
\begin{aligned}
\hat s&=\left\lbrace \sin{\theta}\sin{\phi},\cos{\phi},\cos{\theta}\sin{\phi}\right\rbrace\cr
\end{aligned}
\label{shatdef}
\end{equation}
we have obtained the elements of \Eq{EvecEperp}.  
Using for the time derivatives of a vector the operator $(\partial\psi/\partial t)\partial/\partial \psi$,
and for the square the dot product of the vector with itself, \Eq{Pa1a} was evaluated using standard Mathematica
vector operators, yielding an analytical expression for $\partial P/{\partial\varOmega}$ of considerably higher complexity
than Peters\&Mathews Eq. 14.
The instantaneous total power $P(\psi)$ is then a function of $\psi$, obtained by integration over the sphere: 
\begin{equation}
\begin{aligned}
P(\psi)=\int_{4\pi}\frac{\partial P}{\partial\varOmega} \, d \varOmega
=\int_0^{2\pi}\int_0^\pi \frac{\partial P}{\partial\varOmega} \,\sin{\phi}\, d \phi\,\, d\theta
\end{aligned}
\label{sphereintegral}
\end{equation}
The output of this integration is much less complex because
of the smaller number of variables:
\begin{equation}
P(\psi)=P_{\rm com}\left(\big(1+\epsilon\cos{(\psi )}\big)^2+\frac{\epsilon^2 }{3} \sin^2{(\psi )}\right)
\end{equation}
whereas the corresponding Peters\&Mathews Eq. 15 is:
\begin{equation}
P_{\rm PM}(\psi)=P_{\rm com}\left(\big(1+\epsilon\cos{(\psi )}\big)^2+\frac{\epsilon^2 }{12} \sin^2{(\psi )}\right)
\end{equation}
where, in both cases,
\begin{equation}
P_{\rm com}=\frac{32\, G^4 {m_1}^2{m_2}^2({m_1}+{m_2})\big(1+\epsilon  \cos{(\psi)}\big)^4}{5\, a^5 c^5\left(1-\epsilon^2\right)^5}
\end{equation}
It is quite remarkable that these two approaches, starting with different fundamental theories of gravitation
which predict drastically different radiation patterns, come up
with total radiated power of exactly the same functional form, and differ only in the coefficient of the $\epsilon^2\sin^2{(\psi )}$ term. 
This difference may be detectable in systems of larger eccentricity than those yet studied.

The average power output of the binary is obtained by
integrating the instantaneous power over one orbital period $P_b$:
\begin{equation}
\begin{aligned}
\left< P\right> &=\frac{1}{P_b}\int_0^{2\pi} \frac{P(\psi)}{\partial\psi/\partial t}\, d \psi
=\frac{32 G^4 {m_1}^2 {m_2}^2 ({m_1}+{m_2})}{5 a^5 c^5 \left(1-\epsilon^2\right)^{7/2}}
 \left(1+\frac{19 \epsilon ^2}{6}+\frac{5 \epsilon ^4}{12}\right)
\end{aligned}
\label{psiintegral}
\end{equation}
which is to be compared to Peters\&Mathews Eq. 16:
\begin{equation}
\begin{aligned}
\left< P_{PM}\right> &=\frac{32 G^4 {m_1}^2 {m_2}^2 ({m_1}+{m_2})}{5 a^5 c^5 \left(1-\epsilon^2\right)^{7/2}}
 \left(1+\frac{73 \epsilon ^2}{24}+\frac{37 \epsilon ^4}{96}\right)
\end{aligned}
\label{PandMresult}
\end{equation}

\subsubsection{Time Derivative of Binary Period}

What is measured in observations of distant binary systems is not the energy loss, but the temporal change
in the orbital period $P_b$ of the binary system.   As the system loses energy due to gravitational radiation,
the two masses sink deeper into their mutual potential well.   The total energy $E$ of the system is 
\begin{equation}
\begin{aligned}
E=-\frac{m_1 m_2G}{2a}\quad\Rightarrow\quad \frac{\partial E}{\partial t}&=\frac{m_1 m_2G}{2a^2}\frac{\partial a}{\partial t}
\quad\Rightarrow\quad\frac{1}{E}\frac{\partial E}{\partial t}&=-\frac{1}{a}\frac{\partial a}{\partial t}
\label{Wb}
\end{aligned}
\end{equation}

Taylor and Weisberg\cite{Taylor82} quote the 1963 Peters and Mathews result,  and
also quote Wagoner\cite{Wagoner}, where the period derivative is worked out for Kepler's law.  From \Eq{randdpsidt}:
\begin{equation}
\begin{aligned}
&P_b =2\pi \frac{a^{3/2}}{\sqrt{G\left(m_1+m_2\right)}}
\qquad\Rightarrow\qquad \frac{\partial P_b}{\partial t}=\frac{6\pi}{2} \frac{a^{1/2}}{\sqrt{G\left(m_1+m_2\right)}}\frac{\partial a}{\partial t}
\label{Kepler1}
\end{aligned}
\end{equation}
And therefore
\begin{equation}
\begin{aligned}
\frac{1}{P_b}\frac{\partial P_b}{\partial t}&=\frac{3}{2a}\frac{\partial a}{\partial t}
=-\frac{3}{2E}\frac{\partial E}{\partial t}=-\frac{3a}{m_1m_2G}\frac{\partial E}{\partial t}\cr
&=-\frac{96}{5} \frac{G^3}{c^5}\frac{m_1 m_2\left(m_1+m_2\right)}{a^4 \left(1-\epsilon^2\right)^{7/2}}
 \left\{1+\frac{19 \epsilon ^2}{6}+\frac{5 \epsilon ^4}{12}\right\}
\label{dPdt}
\end{aligned}
\end{equation}
where the final result, which uses $\left<P\right>=-\partial E/\partial t$ from \Eq{psiintegral}, is slightly different from
the GR result (\Eq{PandMresult}) in the coefficients of the higher order $\epsilon$ terms in braces. 

For many astronomical situations, the semi-major axis is not easy to infer, whereas the period $P_b$ is directly observable. 
So a more useful form of \Eq{dPdt} is obtained by eliminating $a$ using \Eq{Kepler1}:
\begin{equation}
\begin{aligned}
\frac{1}{a^{3/2}} &=\frac{2\pi}{P_b}\frac{1}{\sqrt{G\left(m_1+m_2\right)}}\cr
\frac{P_b}{a^{4}} &=\left(\frac{2\pi}{P_b}\right)^{8/3}\frac{P_b}{2\pi }\,\frac{2\pi}{(G\left(m_1+m_2\right))^{4/3}}
=\left(\frac{2\pi}{P_b}\right)^{5/3}\frac{2\pi}{(G\left(m_1+m_2\right))^{4/3}}
\label{dPdt2}
\end{aligned}
\end{equation}
upon which \Eq{dPdt} becomes
\begin{equation}
\begin{aligned}
\dot P_b=\frac{\partial P_b}{\partial t}&=-\frac{192\pi}{5} \left(\frac{G}{c^3}\right)^{5/3}\left(\frac{2\pi}{P_b}\right)^{5/3}
\frac{m_1 m_2}{\left(m_1+m_2\right)^{1/3}}\
 \frac{1+\frac{19 \epsilon ^2}{6}+\frac{5 \epsilon ^4}{12}}{\left(1-\epsilon^2\right)^{7/2}}\qquad{\rm G4v}\cr
\dot P_b=\frac{\partial P_b}{\partial t}&=-\frac{192\pi}{5} \left(\frac{G}{c^3}\right)^{5/3}\left(\frac{2\pi}{P_b}\right)^{5/3}
\frac{m_1 m_2}{\left(m_1+m_2\right)^{1/3}}\
 \frac{1+\frac{73 \epsilon ^2}{24}+\frac{37 \epsilon ^4}{96}}{\left(1-\epsilon^2\right)^{7/2}}\qquad{\rm GR}
\label{dPdt3}
\end{aligned}
\end{equation}
The quantity $\dot P_b$ is one of the standard {\bf post-Keplerian parameters}\cite{DandD}, and is slightly different in
the numerator $\epsilon^2$ term from that obtained with GR (at first post-Newtonian order, 1PN) as noted above.  
The $\epsilon^2$ term in the G4v expression
is 4\% larger than that of the GR expression, making the predicted
energy loss rate for the double pulsar J0737-3039A/B ($\epsilon=0.08777$) larger by only $\approx10^{-3}$ than the GR prediction. 
Unfortunately this difference is too small to be resolved with data currently available\cite{Burgay12}, but may come in range
as observations continue and new systems are discovered.  For the Hulse-Taylor pulsar PSR
B1913+16 ($\epsilon=0.617$) the G4v prediction is $\approx 2\%$ larger than the Peters\&Mathews GR value.  
The measured period derivative for this system is about 1\% higher than the Peters\&Mathews value\cite{Weisberg_04}, 
but is ``corrected'' downward by a poorly-know acceleration relative to the solar system, thus making
it agree with the GR prediction.  At face value this result would place the G4v value outside the error bounds
on the measured period derivative.  Caution is in order, however, not only because of the uncertainty of the correction, but
because both calculations use the classical Kepler model
for the binary, and, at least for G4v, a full model has yet to be developed.  Such a self-consistent model is expected to have
corrections of the order of this apparent disagreement, so the entire question awaits such an analysis.


\subsection{Lumped Pulsar Model}
\begin{figure}[!h]
\begin{center}
\includegraphics{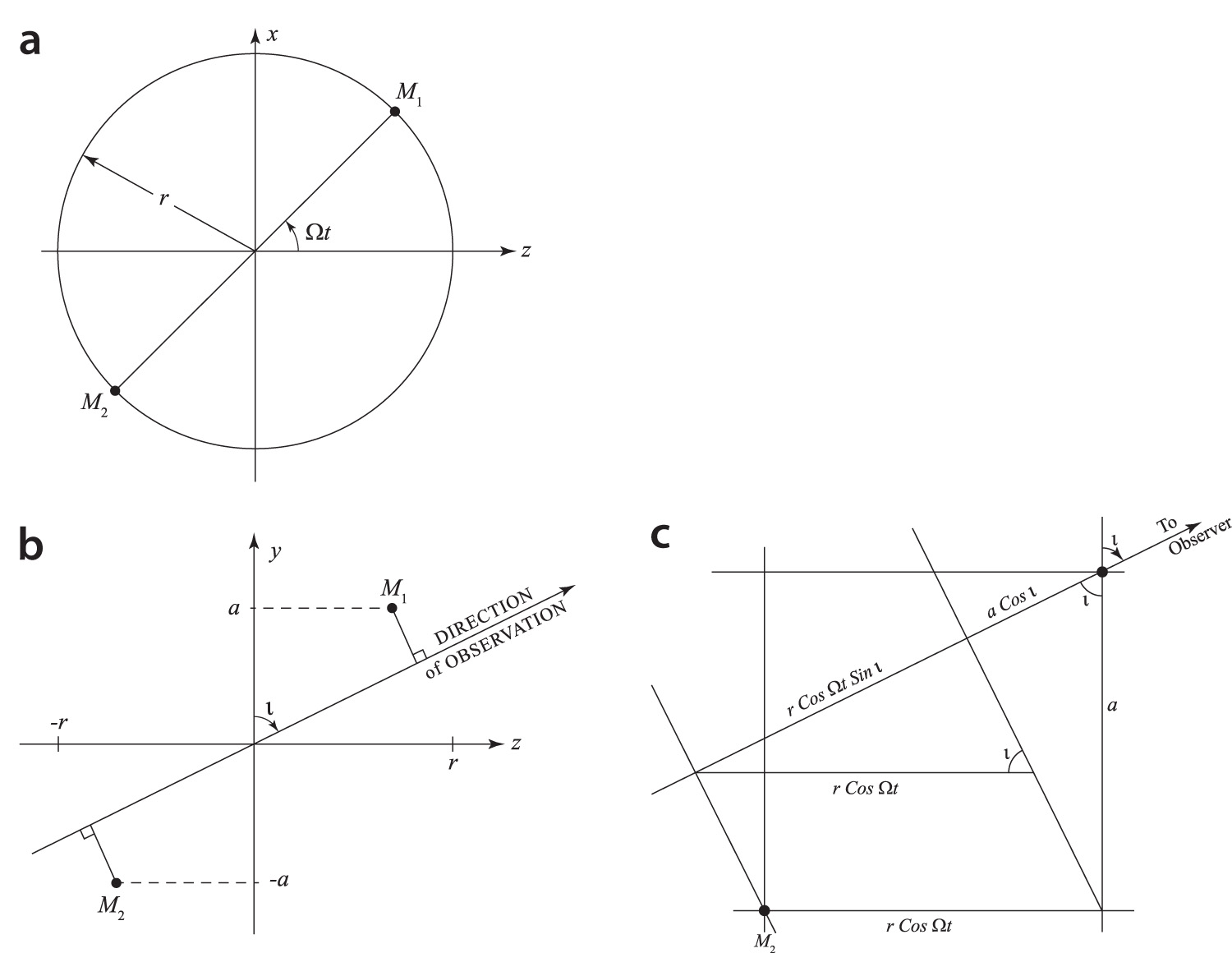} 
\caption{Top ({\bf a}) and side ({\bf b}) views of lumped pulsar gravitational wave source. 
The $y$ axis is the axis of rotation, and the radius of both orbits is $r$.
The mass $M_1$ is located at $y=a$ and $M_2$ is located at $y=-a$ 
Observation is done at a point located at a distance $R\gg r$, in the $y-z$ plane
at an angle $\iota$ from the $y$ axis.  The angle $\iota$ is the {\bf inclination}
of the binary with respect to the line of sight. \hfil\break 
({\bf c}) Magnified third quadrant of side view showing construction for \Eq{Tdt1}.
\label{TBinary}}
\end{center}
\end{figure}

We now consider the somewhat more speculative source of gravitational waves, as shown in \Fig{TBinary}:  a system 
consisting of two equal masses $M_1=M_2=M$ orbiting their common
center of mass with angular frequency $\Omega$, but located in planes of different $y$.

Although such a configuration cannot exist as an isolated binary, Alan Weinstein suggested that
a perturbation of this form may occur as a deformation of a neutron star, and therefore
we should not ignore signals produced by such a configuration in our searches for gravitational waves from pulsars.

We can motivate the conjecture as follows:  Pulsars, by their very nature, have a strong magnetic dipole whose axis is
inclined with respect to the axis of rotation.  The magnetic vector potential may be a non-negligible part of the momentum
of any charged species in the pulsar.  Therefore we must include the magnetic contribution to the momentum as part
of the source of the gravitational vector potential.  As such, an appropriate abstraction of the pulsar as a source of
gravitational radiation might be an axially symmetric momentum distribution (giving no gravitational radiation) plus
a perturbation due to the magnetic momentum.  The latter contribution can be modeled most simply as additional
source masses located where the magnetic axis enters and exits the surface of the pulsar.

Whether real pulsars have sufficient ``lumps'' at the entry and exit points of the magnetic axis is a question well beyond
the scope of the present discussion.  It should be pointed out, however, that the existence and magnitude of ``lumps'' of any kind
on real pulsars is still an open question.  For that reason we shall analyze the consequences of magnetically related
lumps from the observational and energy-loss point of view without further apology.

The positions of the two masses are shown in \Fig{TBinary}:
\begin{equation}\begin{aligned}
z_{1} &=\ r \cos{\Omega t}\qquad\qquad\  y_1=a\cr
z_{2} &=-r \cos{\Omega t}\qquad\qquad y_2=-a\cr
\end{aligned}
\label{Tpos}
\end{equation}
The time difference $\delta t$ along the path to the observer 
is shown in the lower-right of \Fig{TBinary}:
\begin{equation}
\begin{aligned}
c\, \delta t_1 &=-r \cos{\Omega t} \sin{\iota}-a\cos{\iota}\cr
c\, \delta t_2 &=\ r \cos{\Omega t} \sin{\iota}+a\cos{\iota}\cr
\end{aligned}
\label{Tdt1}
\end{equation}

where the reference position ($\delta t=0$) is taken at the center of the binary.  We restrict this simple example to non-relativistic velocities.
The z-directed velocities of $M_1$ and $M_2$ are exactly as given in the binary case:
\begin{equation}
\begin{aligned}
v_{z1} &=\frac{\partial z_1}{\partial t}
=-\Omega r\sin{\Omega t}\cr
v_{z2} &=\frac{\partial z_2}{\partial t}
=\Omega r\sin{\Omega t}\cr
\end{aligned}
\label{Tvz}
\end{equation}
The velocities $v_1$ and $v_2$ with respect to the observer are
those given above multiplied by $\sin{\iota}$, also the same as the binary case.

\subsubsection{Vector Potential} 
The vector potential $\vec A$ at an observation point at distance $R\gg r$
is the sum of the Green's functions of the individual masses
exactly as we do in electromagnetic propagation ({\bf CE} 4.5):
\begin{equation}
\begin{aligned}
{\vec A}={\chi}\sum_i\frac{\vec k_i}{r_i}
=\frac{G\hbar}{c^3}\sum_i\frac{\vec k_i}{r_i}
=\frac{G}{c^3}\sum_i\frac{\vec p_i}{r_i}
\approx\frac{G}{c^3}\sum_i\frac{m_i \vec v_i}{r_i}
\end{aligned}
\label{TGreen4}
\end{equation}

The $x$ components of the momenta are
\begin{equation}\begin{aligned}
p_{x1} \approx M_1v_{1x}&=M\Omega\, r \cos{\Omega t}\cr
p_{x2} \approx M_2v_{2x}&=-M\Omega\, r \cos{\Omega t}\cr
\end{aligned}
\label{Txvel}
\end{equation}
The $z$ components of the momenta are

\begin{equation}\begin{aligned}
p_{z1} \approx M_1v_{1z}&=-M\Omega\, r \sin{\Omega t}\cr
p_{z2} \approx M_2v_{2z}&=M\Omega\, r \sin{\Omega t}\cr
\end{aligned}
\label{Tzvel}
\end{equation}
The time delays, as given by \Eq{Tdt1}, are the only difference between the lumped case
and the symmetric binary case insofar as the far-field radiation properties are concerned.

We calculate the the $x$ and $z$ components of $\vec A$ separately:

\subsubsection{${\bf A}_x$ component}
\noindent
From \Eq{TGreen4} and \Eq{Txvel}
\begin{equation}
A_x(R,t)\approx \frac{G}{c^3}\left(
\ \frac{p_{x1}\left( t-\delta t\right)}{R}
\ +\ \frac{p_{x2}\left( t+\delta t\right)}{R}
\right)
\label{TArtd}
\end{equation}
where we have neglected the contribution of the $x$ location to the separation since $R\gg r$.\\  Using \Eq{Txvel} 
\begin{equation}
\begin{aligned}
A_x(R,t)\approx {\frac{GM\Omega\, r}{R\, c^3}} \Big(
&\cos{\Omega{\left( t-\delta t\right)}}
\ -\ {\cos{\Omega{\left( t+\delta t\right)}}}
\Big)\cr
= {\frac{GM\Omega\, r}{R\, c^3}} \Big(
&\cos{\Omega t}\ \cos{\Omega \delta t}+ \sin{\Omega t}\ \sin{\Omega \delta t}\cr
-&\cos{\Omega t}\ \cos{\Omega \delta t}+ \sin{\Omega t}\ \sin{\Omega \delta t}
\Big)\cr
= {\frac{2GM\Omega\, r}{R\, c^3}} \Big(
&\sin{\Omega t}\ \sin{\Omega \delta t}
\Big)
\label{TArtd1}
\end{aligned}
\end{equation}
Using the value of $\delta t$ from \Eq{Tdt1}

\begin{equation}
\begin{aligned}
A_x(R,t)&\approx \frac{2GM\Omega\, r}{R\, c^3}\left\lbrack
\sin{\Omega t}\ \sin{\left(\frac{\Omega r }{c}
\big(\cos{\Omega t}\,\sin{\iota}+\frac{a}{r}\cos{\iota}\big)\right)}\right\rbrack\cr
\label{TArtd2}
\end{aligned}
\end{equation}

Assuming $\Omega r\ll c$ \Eq{TArtd2} becomes
\begin{equation}
\begin{aligned}
A_x(R,t)&\approx \frac{2GM\Omega\, r}{R\, c^3}\left\lbrack
\sin{\Omega t}\ {\left(\frac{\Omega r }{c}
\big(\cos{\Omega t}\,\sin{\iota}+\frac{a}{r}\cos{\iota}\big)\right)}\right\rbrack\cr
&=\frac{2GM\Omega^2\, r^2}{R\, c^4}\left\lbrack
\sin{\Omega t}\ \big(\cos{\Omega t}\,\sin{\iota}+\frac{a}{r}\cos{\iota}\big)\right\rbrack\cr
&=\frac{GM\Omega^2\, r^2}{R\, c^4}\Big(
\sin{2\Omega t}\, \sin{\iota}+2\frac{a}{r}\sin{\Omega t}\, \cos{\iota}\Big)\cr
\label{TArtd3}
\end{aligned}
\end{equation}
Comparing this result with that of \Eq{Artd2}, we see that the lumped pulsar produces far-field radiation at
the fundamental rotation frequency as well as at twice the rotation frequency.  The amplitudes of the two frequency components depend on the angle $\alpha$ between the spin axis and the magnetic axis since $a=R_0 \cos{\alpha}$ and $r=R_0\sin{\alpha}$.
\begin{equation}
\begin{aligned}
A_x(R,t)&\approx h_0\sin{\alpha}\Big(
\sin{\alpha}\sin{2\Omega t}\, \sin{\iota}+2\cos{\alpha}\sin{\Omega t}\, \cos{\iota}\Big)\cr
&{\rm where}\qquad h_0= \frac{GM\Omega^2\, R_0^2}{R\, c^4}
\label{TArtd3a}
\end{aligned}
\end{equation}
The $x$-directed field ${\cal E}_x$ is given by
\begin{equation}
{\cal E}_x=-\frac{\partial A_x}{\partial t}
\approx -\frac{2GM\Omega^3\, r^2}{R\, c^4}
\Big(\cos{2\Omega t}\ \sin{\iota}+\frac{a}{r}\cos{\Omega t}\ \cos{\iota}\Big)
\label{Tpropsol1}
\end{equation} 
\subsubsection{${\bf A}_\iota$ component}
The Transverse Theorem tells us that only the transverse vector potential contributes to interaction with matter.
In the coordinate system we are using, the components of the transverse vector potential $A_\perp$ are
$A_x$ and $A_\iota$.
\begin{equation}
\begin{aligned}
A_\iota(R,t)=A_z \cos{\iota}&=  \frac{GM\Omega^2\, r^2}{R\, c^4}
 {\left(\cos{2\Omega t}\,\sin{\iota}\cos{\iota}+2\,\frac{a}{r}\cos{\Omega t}\,\cos^2{\iota}\right)}\cr
&=  h_0\sin{\alpha}
 {\left(\sin{\alpha}\cos{2\Omega t}\,\sin{\iota}\cos{\iota}+2\cos{\alpha}\cos{\Omega t}\,\cos^2{\iota}\right)}
\label{TAtheta}
\end{aligned}
\end{equation}
The $\iota$-directed  field ${\cal E}_\iota$ is given by
\begin{equation}
\begin{aligned}
{\cal E}_\iota=-\frac{\partial A_\iota}{\partial t}
&\approx \frac{2GM\Omega^3\, r^2}{R\, c^4}
{\left(\sin{2\Omega t}\,\sin{\iota}\cos{\iota}+\,\frac{a}{r}\sin{\Omega t}\,\cos^2{\iota}\right)}
\label{TEtheta}
\end{aligned}
\end{equation} 

\begin{figure}[!hb]
\begin{center}
\includegraphics{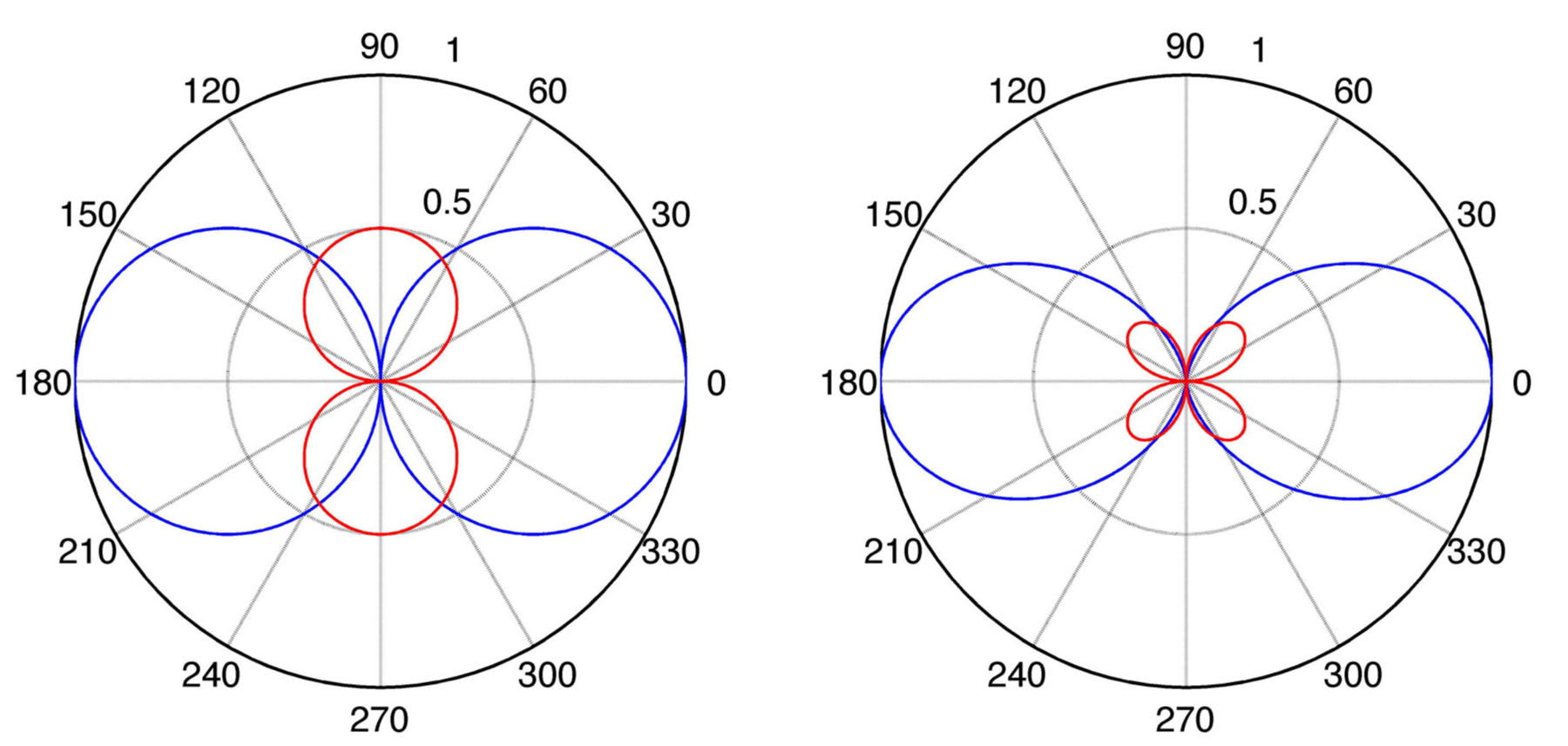} 
\caption{Comparison of radiation patterns of a lumped pulsar with ``lumps'' inclined $45^\circ$ from its rotation axis.
Two polarizations of the transverse vector potential are shown: $A_x$ (left) and $A_\iota$ (right). 
Vector potential amplitudes (as multiples of $h_0$) are plotted in the radial direction
{\it vs} inclination  $\iota$.  The zero of inclination is when the source rotation axis is pointing toward the observer.
Blue is the $\Omega$ amplitude, and  red is the $2\Omega$ amplitude.   
\label{LumpRadPat}} 
\end{center}
\end{figure}

\subsubsection{Radiation Patterns}

The radiation patterns of \Eq{TArtd2} and \Eq{TAtheta} are plotted in \Fig{LumpRadPat} for an arbitrarily chosen
angle of $45^\circ$ between the spin axis and rotation axis.  We see that the $\Omega$ radiation can be large, and
has a maximum directed along the axis of rotation, whereas the $2\Omega$ radiation is always zero along the axis of rotation.

\subsubsection{Total Radiated Power}
As in the symmetric binary case, the total power radiated $P$ is obtained by integrating
Poynting's Vector $S$ from \Eq{Pa} over the sphere of radius $R$
\begin{equation}
\begin{aligned}
P&=\int_{0}^{\pi}S\ 2\pi R \sin{\iota}\ Rd\iota\cr
&=\frac{ R^2 c^3}{2 G}\ \int_{0}^{\pi}
\Big(\left<{\cal E}_x^2\right>+\left<{\cal E}_\iota^2\right>\Big)
\sin{\iota}\ d\iota
\label{TPa1}
\end{aligned}
\end{equation}

From\Eq{Tpropsol1} we have
\begin{equation}
\begin{aligned}
{\cal E}^2_x&\approx \frac{4G^2M^2\Omega^6\, r^4}{R^2\, c^8}
\Big(\cos^2{2\Omega t}\ \sin^2{\iota}+\frac{a^2}{r^2}\cos^2{\Omega t}\ \cos^2{\iota}\Big)
\label{TE2x}
\end{aligned}
\end{equation}
where the cross terms average to zero due to orthogonality of  $\Omega t$ and $2\Omega t$.

From\Eq{TEtheta} we have
\begin{equation}
\begin{aligned}
{\cal E}^2_\iota&=\frac{4G^2M^2\Omega^6\, r^4}{R^2\, c^8}
{\left(\sin^2{2\Omega t}\,\sin^2{\iota}\,\,\cos^2{\iota}+\,\frac{a^2}{r^2}\sin^2{\Omega t}\,\cos^4{\iota}\right)}\cr
&=\frac{4G^2M^2\Omega^6\, r^4}{R^2\, c^8}
{\left(\sin^2{2\Omega t}\,\Big(\sin^2{\iota}-\sin^4{\iota}\Big)
+\,\frac{a^2}{r^2}\sin^2{\Omega t}\,\cos^4{\iota}\right)}
\label{TE2theta}
\end{aligned} 
\end{equation} 

Because $\left<\cos^2{\Omega t}\right>=\left<\sin^2{\Omega t}\right>=
\left<\cos^2{2\Omega t}\right>=\left<\sin^2{2\Omega t}\right>=1/2$,\\ 
\Eq{TPa1} for the average power becomes
\begin{equation}
\begin{aligned}
P&\approx{\frac{GM^2\Omega^6\, r^4}{c^5}}\int_0^{\pi}
\left(2\sin^2{\iota}-\sin^4{\iota}
+\,\frac{a^2}{r^2}\big(\cos^2{\iota}+\cos^4{\iota}\big)\right)\sin{\iota}\ d\iota \cr
&={\frac{GM^2\Omega^6\, r^4}{c^5}}\left(
\int_0^{\pi}\Big(2\sin^3{\iota}-\sin^5{\iota}\Big)\ d\iota 
+\frac{a^2}{r^2}\int_0^{\pi}\big(\cos^2{\iota}+\cos^4{\iota}\big)\sin{\iota}\ d\iota \right)\cr
&={\frac{GM^2\Omega^6\, r^4}{c^5}}\left(\frac{8}{5} +\frac{16}{15}\frac{a^2}{r^2}\right)
\label{TP2}
\end{aligned}
\end{equation}
The first term, from the $2\Omega$ radiation, is the same as \Eq{P2}, and reduces to it when the masses are in the same plane.
The second term is from the $\Omega$ radiation, which has a quite different radiation pattern from that of the $2\Omega$ radiation.
We can evaluate the relative contributions of the two components by recognizing that $r^2+a^2=R_0^2$, where $R_0$ is the radius of the pulsar,
and $a/R0=\cos{\alpha}$:

\begin{equation}
\begin{aligned}
P&\approx{\frac{GM^2\Omega^6}{c^5}}\left(R_0^2-a^2\right)\left(\frac{8}{5}\left(R_0^2-a^2\right) +\frac{16}{15}{a^2}\right)\cr
&\approx{\frac{GM^2\Omega^6R_0^4}{c^5}}\left(1-\frac{a^2}{R_0^2}\right)
\left[\frac{8}{5}\left(1-\frac{a^2}{R_0^2}\right) +\frac{16}{15}\frac{a^2}{R_0^2}\right]\cr
&\approx{\frac{GM^2\Omega^6R_0^4}{c^5}}\sin^2{\alpha}
\left[\frac{8}{5}\sin^2{\alpha} +\frac{16}{15}\cos^2{\alpha}\right]
\label{TP2a}
\end{aligned}
\end{equation}

\begin{figure}[!ht]
\begin{center}
\includegraphics[height=7cm]{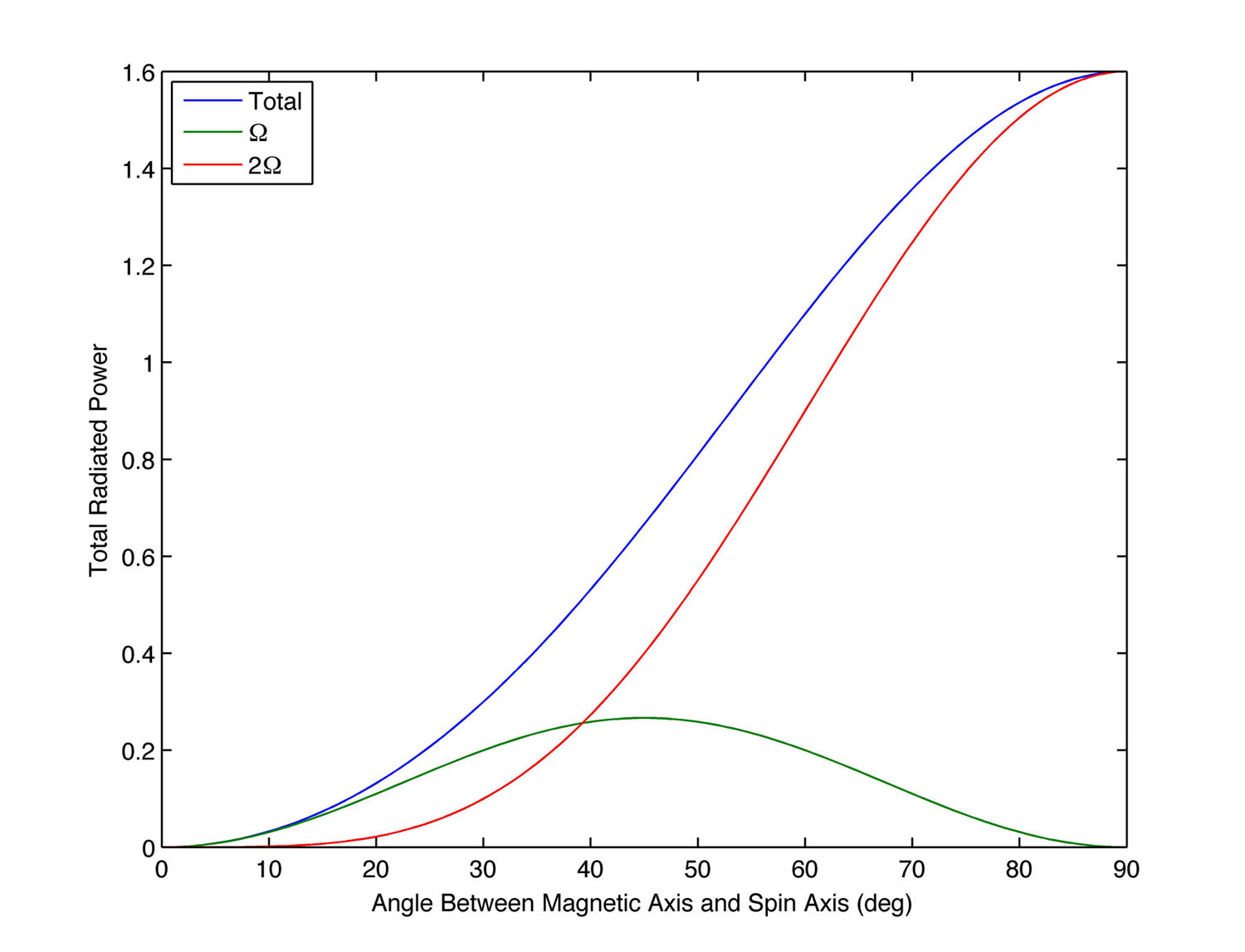} 
\caption{Power radiated from a lumped pulsar as a function of the angle of misalignment between rotation axis and magnetic axis. 
At large angles, radiation is primarily at the second harmonic of the rotation frequency $\Omega$.  At angles less than $\approx 40^\circ$,
radiation at the fundamental frequency dominates.
\label{LumpTotPwr}}
\end{center}
\end{figure}

The power radiated by the two frequency components is shown in \Fig{LumpTotPwr}.  It should be noted that the $\Omega$
radiation only makes a contribution larger than the $2\Omega$ radiation when the angle between the magnetic axis
and the spin axis is less than $\approx 40^\circ$.  In this region, the pulsar is not a very efficient radiator.

\subsubsection{Discussion of the Results}
Several aspects of the present treatment deserve comment:

 All derivations have been done directly with the Green's functions 
of the source elements, so we have never had to resort to any kind of multipole expansion.  The $2\Omega$ radiation carries
the same energy as that predicted by Peters and Mathews for their quadrupole source, and we have every reason to refer
to it in those terms as well.  In this structure, the energy radiated by the 
$\Omega$ component is always less than that radiated at $2\Omega$ by a binary system formed from the same two perturbing masses.
It seems that something like the lumped pulsar is conceptually possible, so it is important to include the fundamental frequency
in coherent searches for gravitational waves from known pulsars.  Similar conclusions were reached recently by Bejger \&
Kr'olak\cite{Bejger}.  However it may turn out that no $1\Omega$ radiation is possible from an isolated structure for some
symmetry/conservation reason.  The resolution of this question remains in the category of future work.

\section{Gravitational Wave Detection}
There are three distinct ways in which a gravitational wave of this type can interact with a detector:
\begin{enumerate}
\item{The transverse vector potential $\vec A_\perp$ appears directly in the momentum of a mass in the detector,
as discussed in the preceding sections.}
\item{The scalar potential directly modulates the speed of light in the detector.}
\item{The vector potential appears directly as part of the vector velocity of light in the detector.}
\end{enumerate}
In the long-wavelength limit, effects 2 and 3 are of higher order than effect 1 by
a factor of order $\Omega l/c$.  We therefore limit the present discussion to the direct effect of the transverse
vector potential on the positions of the two detector masses.

\subsection{Motion of Detector Masses}
From \Eq{propsol}, and \Eq{Atheta}, the two components of the propagating transverse vector potential $A_\perp$ from a binary source are
 \begin{equation}
\begin{aligned}
A_x&\approx h_0\ \sin{2\Omega t}\ \sin{\iota}\qquad{\rm where}\qquad h_0= \frac{GM\Omega^2\, r^2}{R\, c^4}\cr
A_\iota&\approx h_0\  \cos{2\Omega t}\ \sin{\iota} \cos{\iota}
\label{Potentials}
\end{aligned}
\end{equation}
In the gravitational-wave literature, $h_0$ is called the {\bf dimensionless strain}.

\noindent
As discussed in Section \ref{MassMotion},
the vector potential is the direct measure of how the local frame of reference is affected 
by the movement of distant matter.  The fractional change $\delta l$ in length $l$ between the two free-floating masses given by \Eq{dl} is
\begin{equation}
\begin{aligned}
\frac{\delta l}{l}&\approx
-\left(\vec A\cdot{\hat l}\right)\left(\hat R\cdot\hat l\right)
\label{dl1}
\end{aligned}
\end{equation}
where $\hat R$ is a unit vector in the direction from the source to the detector.\\
We can understand this formula intuitively in the following way:\\
The first term arises because the velocity of each individual mass is in the direction of $\vec A$.
The second term expresses the fact that the position of each mass is the velocity times the time.
Thus the difference in the positions will be the difference in vector potential times the difference in time of arrival.
\subsection{Detector Sensitivity}
\begin{figure}[!h]
\begin{center}
\includegraphics{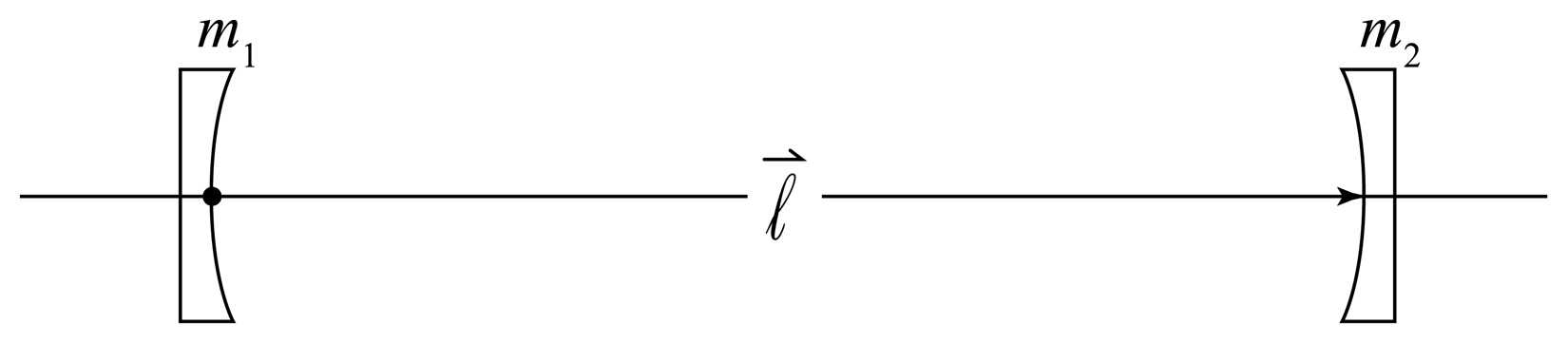} 
\caption{Side view of Fabry-P\'erot cavity used in gravitational wave detector. 
The vector $\vec l$ is the length of the cavity along the axis of light propagation.} 
\label{cavity}
\end{center}
\end{figure}

In today's world, credible gravitational-wave detectors use laser interferometers whose
massive mirrors are suspended from highly compliant springs.  This mounting arrangement
is necessitated by the need to isolate the mirrors from vibrations of the earth.  The axis of
the interferometer is parallel to the earth's surface, so, for small movements along the axis of
light propagation, the mounts are virtually free-floating---they experience essentially no restoring force.
Each arm of the interferometer contains a Fabry-P\'erot cavity, a single example of which is shown 
in \Fig{cavity}.

We have already analyzed the effect of the transverse vector potential on $\vec l$,
the vector length of the cavity.  The number of fringes $n$ of light of wavelength $\lambda$ 
in a cavity resonator is $n=2l/\lambda=2lc/f$.  

Using \Eq{dl1} for $\delta l/l$ we obtain
\begin{equation}
\begin{aligned}
\frac{\delta n}{n}&\approx 
-\left(\vec A_\perp \cdot{\hat l}\right)\left(\hat R\cdot\hat l\right)
\label{dntot}
\end{aligned}
\end{equation}
So, to evaluate the fringe shift due to a gravitational wave, only two projection operations are needed
for each polarization of $A_\perp$:
The projection of the transverse vector potential onto the centerline of the cavity, and the projection
of the centerline of the cavity onto the direction of observation.  

\begin{figure}[!h]
\begin{center}
\includegraphics{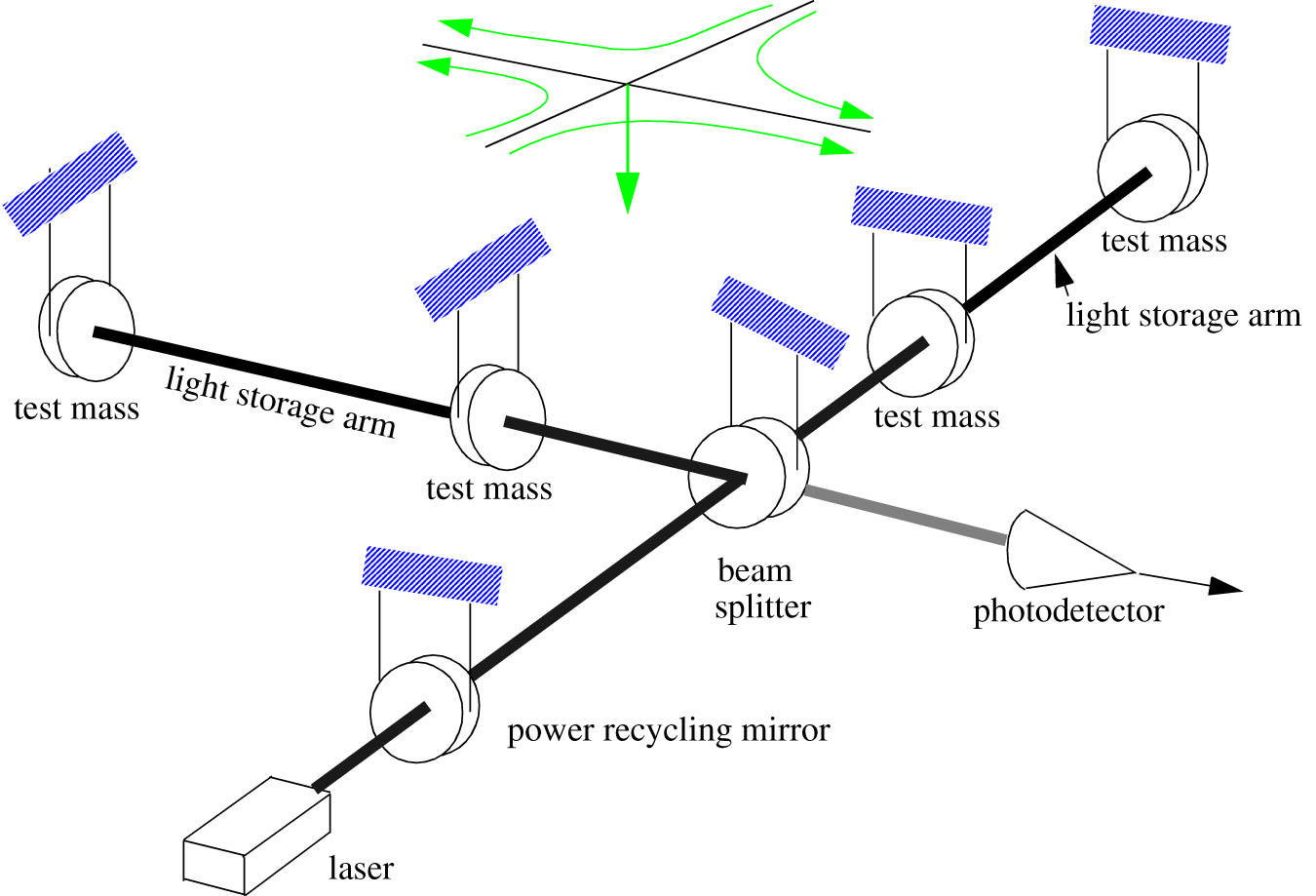} 
\caption{Cartoon of LIGO interferometric gravitational wave detector, which has two Fabry-P\'erot cavities, labeled
``light storage arm'' in this drawing.  The output of the photodetector is proportional to the difference between the phases of the two arms,
and hence is sensitive to a signal that ``stretches'' the distance between the mirrors in one arm and ``squeezes''
the distance between the mirrors in the other.  From the LIGO web site\cite{Barish99}.} 
\label{LIGO_Dwg2}
\end{center}
\end{figure}

The actual observatory instrument uses two cavities at right angles to each other, as shown in \Fig{LIGO_Dwg2}.

Because the interferometer takes the difference between the fringe shifts of the two arms, 
the sensitivity of the detector to the two polarizations is, from \Eq{dntot}
\begin{equation}
\begin{aligned}
S_x&=\left(\hat l_1\cdot \hat A_x\right)\left(\hat l_1\cdot \hat R\right)
-\left(\hat l_2\cdot \hat A_x\right)\left(\hat l_2\cdot \hat R\right)\cr
S_\iota&=\left(\hat l_1\cdot \hat A_\iota\right)\left(\hat l_1\cdot \hat R\right)
-\left(\hat l_2\cdot \hat A_\iota\right)\left(\hat l_2\cdot \hat R\right)
\label{SxSy}
\end{aligned}
\end{equation}
These sensitivities are the response of the detector to a vector potential of unit amplitude in the 
$x$ and $\iota$ directions of \Fig{Binary}.  The actual fringe shift due to a signal from a distant source
is obtained by using the actual vector potentials $A_x$ and $A_\iota$, given by \Eq{Potentials}.
These two vector components of the total potential $\vec A$ are called the {\bf polarizations} of the wave.
The output $V$ of the LIGO detector is the sum of the two polarizations, each multiplied by its respective sensitivity:
 \begin{equation}
\begin{aligned}
V=A_x S_x+A_\iota S_\iota \approx h_0\Big(S_x \sin{2\Omega t}\ \sin{\iota}+
S_\iota  \cos{2\Omega t}\ \sin{\iota} \cos{\iota}\Big)
\label{Potentials2}
\end{aligned}
\end{equation}

\begin{figure}[!h]
\begin{center}
\includegraphics{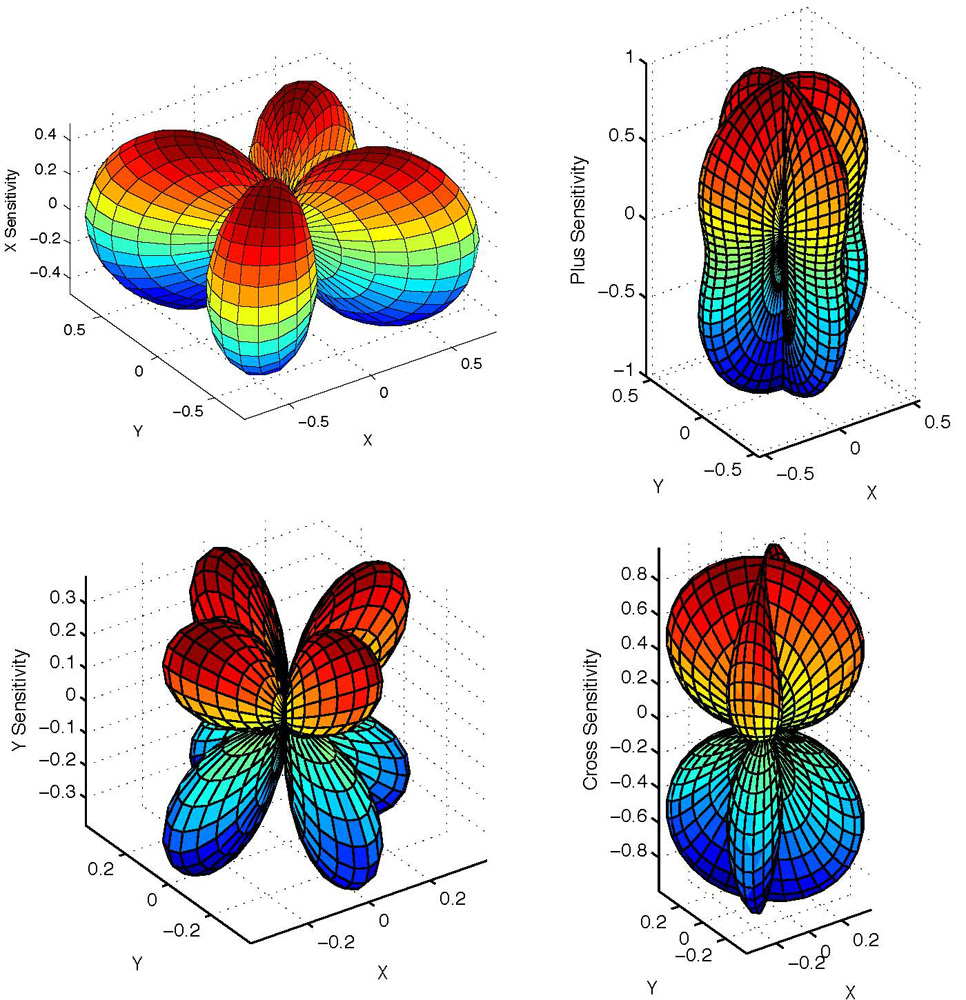}
\caption{Antenna patterns for the two polarizations of G4v (left) and GR (right).
The observatory antenna arms lie along the $x$ and $y$ axes.  The photodetector output due to a gravitational wave of unit $h_0$
is plotted as the radial distance from the origin to the surface as a function of the direction of incoming wave propagation.
\label{fig:AntPat}}
\end{center}
\end{figure}

\subsection{G4v {\it vs} GR Antenna Patterns}
\label{AntPat}
The sensitivity of a LIGO detector for the two polarizations of G4v and GR are compared in \Fig{fig:AntPat}. 
As the Earth rotates, these antenna patterns sweep over the sky and determine the amplitude of the received signals, 
including sign changes. 

It can be seen by inspection that, if the position of the source in the sky is known, the response
of a detector to the G4v signal will be very different from that of a GR signal.  This topic is analyzed in
detail by Isi {\it et.al.}\cite{Isi14}.

\subsection{General Nature of the Interaction} 
It is interesting that the G4v patterns are almost complementary to the GR patterns.
In particular, the G4v patterns have no sensitivity on the $z$ axis (directly above the detector), 
where both GR patterns have maximum sensitivity.   We can understand the basic
nature of the G4v gravitational wave interaction with matter in the following manner:
The change in length $\delta l$ between the two masses separated by a vector distance $\vec l$ 
due to a gravitational vector potential $\vec A$ is, from \Eq{dl1}
\begin{equation}
\begin{aligned}
\delta l&\approx
-\left(\vec A\cdot{\vec l\,}\right)\left(\hat R\cdot\hat l\right)
\label{dll}
\end{aligned}
\end{equation}
where $\hat R$ is a unit vector in the direction of wave propagation.

The first term arises because the velocity of each individual mass is in the direction of $\vec A$. 
In the far-field, the only components of the vector potential that survive are perpendicular to the direction of propagation. 
Thus, in terms of its effect on matter, the wave is a ``shear wave'', where all elements of matter in a plane perpendicular
to the direction of propagation move in elliptical orbits {\it within that plane}.  Thus elements of matter will remain at constant
spacing in any plane perpendicular to the direction of propagation.

The second term expresses the fact that the position of each mass is the velocity times the time,
so the time for a particular element of matter to be displaced is different due to the difference in
propagation time along the direction of propagation.  Because the motion is perpendicular to the direction of propagation,
elements of matter will remain at constant spacing (to first order) along the direction of propagation.

Thus the difference in the positions will be maximum at $45^\circ$ to the direction of propagation. 
So, at any given time and distance along the path of propagation,
elements of matter will be closer together along one $45^\circ$ direction, and farther apart in the orthogonal $45^\circ$ direction.
Thus, within any plane at $45^\circ$ to the direction of propagation, matter will be alternatively  ``stretched'' along one axis and
''squeezed'' along the orthogonal axis. 
This ``quadrupole'' pattern is similar to that predicted by GR, but rotated from the axis of propagation by $45^\circ$. 
Unless the absolute position and polarization phase of the gravitational source is known, the two theories cannot be distinguished
by their instantaneous effect on ``freely floating'' matter.  

Because observatories like LIGO have two arms perpendicular to each other,
whose displacements are subtracted, the antenna patterns shown in Section~\ref{AntPat} are exactly the same as
the ``stretch-squeeze'' of matter along the corresponding orthogonal axes.  Hence these antenna patterns
have four lobes, as appropriate for a quadrupole gravitational wave.
 
As we have shown, a propagating transverse vector potential produces ``stretch-squeeze'' motion
in free-floating matter much as a tensor wave would, but in planes at $45^\circ$ to the direction of propagation.
The foregoing development is completely parallel to the corresponding derivation for electromagnetic
waves given in [{\bf CE}].   It has been known for many years that electromagnetic waves,
which are vector in nature, propagate perfectly well from quadrupole sources\cite{LLQW}. 

Binary pulsar systems are, in principle, capable of exquisitely precise tests of any theory of gravitation.  Such tests rely on
equally precise models of the binary system using the theory being tested.  Such models are presently being evolved using GR,
and we hope that future versions will include G4v as well.

\subsubsection{Field Quantization}

It should be noted that there is a tendency to identify quadrupole interaction with tensor propagation. 
That identity is only appropriate for theories with quantized radiation fields.  In G4v and CE,
the electromagnetic and gravitational fields are four-vector in character, and neither is quantized. 
In fact, there is not a single shred of evidence for quantization of either the electromagnetic or the gravitational radiation field. 
The Casimir effect is often pointed to as ``proving'' the existence of virtual pairs, and therefore of field quantization.
It was shown by Jaffe\cite{Jaffe05}  that the ``Casimir forces can be computed without reference to zero point
energies.  They are relativistic, quantum forces between charges and currents.''  Virtual pairs thus have
the role of simplifying certain calculations rather than providing evidence for field quantization. 
It is easy to see that virtual pairs, and therefore field quantization of the {\it pro forma} kind that
is routinely taught in contemporary physics classes, cannot be physically real.  If they were,
they would contribute a factor of $10^{120}$ more to the energy density of the universe than
can be estimated from any method of observation.  
See\cite{cosconstant02} for a thoughtful review of this well-known {\bf cosmological constant problem}.
Thus the fictitious energy of vacuum fluctuations is seen as a deep {\it conceptual problem} in prevailing theory
rather than a numerical or ``fine tuning'' problem, as it is often portrayed.

In CE and G4v, the degrees of freedom of matter are not separate from the degrees of freedom of the field.
Quantization occurs in the region where field waves interact with matter waves; not in the field in the absence of matter. 
In this view of the relationship of matter and field, there are no vacuum fluctuations, and the whole
cosmological constant problem does not arise.

\vfil\eject
\section{Sky Searches} 
As the Earth rotates, the antenna patterns of \Fig{fig:AntPat} sweep over the sky and modulate the amplitude of the received signals, 
including sign changes.  Coherent searches for pulsar signals heterodyne the received LIGO signal
against a sine and cosine reference locked to the pulsar radio signal (see Garching\cite{Garching} and LIGO\cite{Pitkin} papers
for details).  The assumed amplitude response patterns are then used as ``matched filter''
templates on the heterodyned signal to bring it up out of the noise.  Often these searches
integrate the signal, multiplied by the template, for a year or more. 

To interpret a signal from a particular source at a particular observatory, we need the sensitivity as a function of 
the position of the source and the phase of Earth rotation.

\subsection{Detector Coordinates}
The detector has a latitude $\Phi$ (from equator), east longitude $\lambda$, 
and $\psi_d$ is the angle between the $y$ arm of the interferometer and the local northerly direction.  For example 
Technical Note LIGO-T980044-08 gives the directions of the x and y arms for the LIGO Hanford and Livingston observatories.  
The orientation for Hanford corresponds to a right-hand rotation from
$y={\rm north},\ x={\rm east}$ by $\psi_d = 180-54=126^\circ$, and that for Livingston to
a right-hand rotation from $y={\rm north},\ x={\rm east}$ by $\psi_d = 180+17.7165=197.7165^\circ$.

\subsection{Source Coordinates}
The source coordinates are its declination $\alpha$ (from equator), its right ascension $\beta$,
its polarization angle $\psi$ and its inclination, shown as $\iota$ in \Fig{Binary}.  
The time dependence has been combined in the local {\bf Sky Angle} $\eta$ of the source from the observatory.
The sky angle advances from to east to west with time.  
A source with larger right ascension $\beta$ appears farther to the east, and thus has a smaller sky angle
when viewed from a given point on Earth.  The same source appears farther to the west and therefore
has a larger sky angle when viewed from a point of larger east longitude $\lambda$.
Therefore $\eta(t) = \lambda- \beta  +\Omega_E t$, where
$\Omega_E$ is the earth rotation rate and $t$ is the local sidereal time. 

A rotating source such as the binary shown in \Fig{Binary} has two additional angles:
\begin{enumerate}
 \item The {\bf Inclination Angle} $\iota$ is the angle from the line of sight
to the rotation axis of the source.   A source with $\iota=90^\circ$ has its rotation axis perpendicular
to the line of sight.  A source with $\iota=0^\circ$ has its rotation axis pointed at the Earth.  
\item The {\bf Polarization Angle} $\psi$ is the left-hand rotation angle about the line of sight
from Earth to the source by which the source rotation axis appears to be rotated from the plane of celestial North.
\end{enumerate}

\subsection{Coordinate Transformations} 
\label{antpat}
To transform the detector coordinates into the source coordinate system,
we first rotate around the z (vertical) axis by an angle  $-\psi_d$ so the $x$ axis
is pointing due east, and the $y$ axis due north.  We then rotate around the x (east) axis 
by the observatory latitude  $\Phi$ to align the north-pointing $y$ axis with the Earth's polar axis.
Next we rotate around the $y$ (now polar) axis by an angle $-\eta$ (negative because a right-hand rotation
goes from west to east) to take out the Earth's rotation and place the $z$ axis in the meridian of the star.
Then we rotate around the $x$ (east) axis by an angle $-\alpha$ to point the $z$ axis at the star.
Then we rotate around the $z$ axis by an angle $-\psi$ to align the $y$ axis with 
polarization angle of the star.  Finally we rotate about the $y$ axis by $\pi$ to transform into the source
frame, where the $z$ axis is identified as the propagation vector $\hat R$, pointing at the Earth, 
and the $y$ axis is in the plane of the source rotation axis, as shown in \Fig{Binary}.  
After these rotations, the result is a vector containing the
projection of the arm unit vector onto $\hat A_x$, $\hat A_\iota$, and $\hat R$ directions in the
source frame of reference.

\begin{equation}
\begin{aligned}
\left[\begin{array}{c}
\hat l_1\cdot \hat A_x\\ \hat l_1\cdot \hat A_\iota\\ \hat l_1\cdot \hat R
\end{array}\right]
&=R_y(\pi)R_z(-\psi)R_x(-\alpha)R_y(-\eta)R_x(\Phi)R_z(-\psi_d)
\left[\begin{array}{c}1\\ 0\\ 0 \end{array}\right]
\label{l1dot}
\end{aligned}
\end{equation}
\begin{equation}
\begin{aligned}
\left[\begin{array}{c}
\hat l_2\cdot \hat A_x\\ \hat l_2\cdot \hat A_\iota\\ \hat l_2\cdot \hat R
\end{array}\right]
&=R_y(\pi)R_z(-\psi)R_x(-\alpha)R_y(-\eta)R_x(\Phi)R_z(-\psi_d)
\left[\begin{array}{c}0\\ 1\\ 0 \end{array}\right]
\label{l2dot}
\end{aligned}
\end{equation}
where the brackets indicate vectors, the $R$'s are rotation matrices about the respective axes
by the angle indicated, and evaluation proceeds from right to left.  

Note that these $R$ rotation matrices rotate the coordinate system (alias rotation), 
not the vector in a fixed coordinate system (alibi rotation), 
and are thus the transpose of the usual alibi rotation matrices.

Because the interferometer takes the difference between the fringe shifts of the two arms, 
the sensitivity of the detector to the two polarizations is, from \Eq{dntot}
\begin{equation}
\begin{aligned}
S_x&=\left(\hat l_1\cdot \hat A_x\right)\left(\hat l_1\cdot \hat R\right)
-\left(\hat l_2\cdot \hat A_x\right)\left(\hat l_2\cdot \hat R\right)\cr
S_\iota&=\left(\hat l_1\cdot \hat A_\iota\right)\left(\hat l_1\cdot \hat R\right)
-\left(\hat l_2\cdot \hat A_\iota\right)\left(\hat l_2\cdot \hat R\right)
\label{SxSy1}
\end{aligned}
\end{equation}
These sensitivities are the response of the detector to a vector potential of unit amplitude in the 
$x$ and $\iota$ directions of \Fig{Binary}.  The actual fringe shift due to a signal from a distant source
is obtained by using the actual vector potentials $A_x$ and $A_\iota$, given by \Eq{Potentials}.
These two vector components of the total potential $\vec A$ are called the {\bf polarizations} of the wave.
The output $V$ of the LIGO detector is the sum of the two polarizations, each multiplied by its respective sensitivity:
 \begin{equation}
\begin{aligned}
V=A_x S_x+A_\iota S_\iota \approx h_0\Big(S_x \sin{2\Omega t}\ \sin{\iota}+
S_\iota  \cos{2\Omega t}\ \sin{\iota}\, \cos{\iota}\Big)
\label{Potentials2a}
\end{aligned}
\end{equation}
In general the actual data will be a series of discrete samples $V_i$ recorded at times $T_i$.
Since the source inclination $\iota$ is not a function of time over the periods involved in the search,
we write \Eq{Potentials2} for $V_i$ as a function of $T_i$
\begin{equation}
\begin{aligned}
V_i=h_0 \Big( a_i\sin{2\Omega T_i}\ +b_i  \cos{2\Omega T_i}\Big)
\label{Potentials3}
\end{aligned}
\end{equation}
where  $a_i=S_x(T_i)  \sin{\iota}$ and $b_i=S_\iota(T_i)  \sin{\iota}\, \cos{\iota}$.\\
If $a_i=b_i$, \Eq{Potentials3} would represent a {\bf circularly polarized} wave.
Because, as shown in \Fig{GRvsG4vRad}, $b_i<a_i$ always, the G4v signal is {\bf elliptically polarized}.

\section{Continuous-Wave Searches}
Any spinning mass distribution that is not uniformly distributed around its rotation axis will radiate
gravitational waves of the kind described, and will therefore be potentially detectable by a LIGO-type detector.
The most promising candidates are those that are the closest to us, have the most ``lopsidedness'',
and about which we have some knowledge of the rotation rate $\Omega$.  A class of objects that meet these
criteria are the pulsars that have resulted from recent cataclysmic astronomical events.  The reception of 
electromagnetic pulses from these objects makes it possible to construct a time base that is presumably
related to the spin frequency, and therefore to the gravitational-wave frequency $\Omega$.

The approach to detecting gravitational waves from these objects was pioneered by the Garching group\cite{Garching}
and has been refined and extended by members of the LIGO and Virgo collaboration.  
See in particular Matt Pitkin's Thesis\cite{Pitkin} and references therein.

The technique starts with a heterodyne (multiplication) of the
original data with both a sine and a cosine function, with
frequency $\omega$ chosen to lie at twice the pulsar rotation rate.
These two heterodyned series are then lowpass
filtered (for reasons of antialiasing) before resampling.
We will refer to the real functions resulting from the
heterodyning of the original data $V_i$ as the cosine and sine quadratures:
\begin{equation}
\begin{aligned}
c_i =V_i \cos{\omega T_i}&=h_0 \Big(a_i\sin{2\Omega T_i}+b_i  \cos{2\Omega T_i}\cr
&= h_0\frac{a_i}{2}\Big(\sin{(2\Omega+\omega) T_i}+\sin{(2\Omega-\omega) T_i}\Big)\cr
&+h_0\frac{b_i}{2}  \Big(\cos{(2\Omega+\omega) T_i}+\cos{(2\Omega-\omega) T_i}\Big)\cr
s_i =V_i \sin{\omega T_i}&=h_0\Big(a_i\sin{2\Omega T_i}+b_i  \cos{2\Omega T_i}\cr
&=h_0\frac{a_i}{2}  \Big(\cos{(2\Omega-\omega) T_i}-\cos{(2\Omega+\omega) T_i}\Big) \cr
&+h_0\frac{b_i}{2}\Big(\sin{(2\Omega+\omega) T_i}-\sin{(2\Omega-\omega) T_i}\Big)
\label{CqSq}
\end{aligned}
\end{equation}
The low-pass filter step removes the sum-frequency terms, leaving
 \begin{equation}
\begin{aligned}
c_i &=h_0\frac{a_i}{2}\sin{(2\Omega-\omega) T_i}+h_0\frac{b_i}{2}\cos{(2\Omega-\omega) T_i}\cr
s_i &=h_0\frac{a_i}{2}\cos{(2\Omega-\omega) T_i} -h_0\frac{b_i}{2}\sin{(2\Omega-\omega) T_i}
\label{CqSqf1}
\end{aligned}
\end{equation}
The radio astronomy community has developed exquisitely precise methods for synchronizing 
time coordinates.  Applying those techniques to the pulsar problem has made possible the creation
of precise sine and cosine waveforms that are locked to the electromagnetic ``beeps'' received from the pulsar.

We thus have at our disposal during the heterodyne step of \Eq{CqSq} a time coordinate such that
 $\omega T_i=2\Omega T_i - \phi $, where the phase $\phi$ between the electromagnetic ``beeps''
 and the gravitational wave is not known.  

\subsection{Search Techniques}
The $i$th sample output of the heterodyne process contains a cosine part $c_i$ and a sine part $s_i$
\begin{equation}
\begin{aligned}
\frac{c_i}{h_0} &=\frac{a_i}{2}\sin{\phi}+\frac{b_i}{2}\cos{\phi}\qquad\qquad
\frac{s_i}{h_0} &=\frac{a_i}{2}\cos{\phi} -\frac{b_i}{2}\sin{\phi}
\label{CqSqfp}
\end{aligned}
\end{equation}
where $a_i$ and $b_i$, given by \Eq{Potentials3}, are periodic functions of the time $T_i$
with period 1 sidereal day.  

The most straightforward technique for recovering a signal from random noise is the {\bf matched filter},
in which we multiply each signal sample by a {\bf template} representing the sensitivity for a unit-amplitude
version of the expected signal at the time $T_i$ of that sample, sum the results over all data samples available,
and divide by the number of samples.   The result is an estimate of the amplitude of the expected signal
actually present in the data.  
The two signals we have are $c_i$ and $s_i$, which we can combine as an analytic function
\begin{equation}
\begin{aligned}
v_i&=\Big(c_i+i s_i\Big)
=\frac{h_0}{2}\Big[{a_i}\sin{\phi}+{b_i}\cos{\phi}+i\big({a_i}\cos{\phi} -{b_i}\sin{\phi}\big)\Big]
\label{Alf1}
\end{aligned}
\end{equation}
If we set  
\begin{equation}
\begin{aligned}
\frac{b_i}{\sqrt{a_i^2+b_i^2}}=\cos{\alpha_i}\qquad{\rm and} \qquad\frac{a_i}{\sqrt{a_i^2+b_i^2}}=-\sin{\alpha_i}
\label{Alf2}
\end{aligned}
\end{equation}
\Eq{Alf1} becomes\footnote{This result could have been obtained immediately by treating both the radiated signal
and the antenna patterns as analytic functions from the beginning.}
\begin{equation}
\begin{aligned}
v_i&=\frac{h_0\sqrt{a_i^2+b_i^2}}{2}\Big[-\sin{\alpha_i}\sin{\phi}+\cos{\alpha_i}\cos{\phi}
-i\big(\sin{\alpha_i}\cos{\phi} +\cos{\alpha_i}\sin{\phi}\big)\Big]\cr
&=\frac{h_0\sqrt{a_i^2+b_i^2}}{2}\Big[\cos{(\alpha_i+\phi)}-i\sin{(\alpha_i+\phi)}\Big]
=\frac{h_0\sqrt{a_i^2+b_i^2}}{2}e^{-i(\alpha_i+\phi)}
\label{Alf3}
\end{aligned}
\end{equation}
and our matched-filter template ${\rm F}_i$ is
\begin{equation}
\begin{aligned}
{\rm F}_i&=\frac{\sqrt{a_i^2+b_i^2}}{2}\ e^{-i(\alpha_i+\phi)}=\frac{\sqrt{a_i^2+b_i^2} }{2}\ e^{-i\alpha_i}\ e^{-i\phi}
\label{MF}
\end{aligned}
\end{equation}

An estimate of the amplitude of the signal can be formed by taking the dot product of $v_i$ and $F_i$,
treated as vectors.  A source with the opposite rotation will reverse the sign of the imaginary part of the signal.
For that reason, a match was computed for both right-hand (RH) and left-hand (LH) rotation:
\begin{equation}
\begin{aligned}
h(\phi)&=\frac{1}{N}\sum_{i=1}^N  \Big(\Re{(v_i)}\Re{( {\rm F}_i)}+\Im{(v_i)}\Im{( {\rm F}_i)}\Big)\qquad{\rm RH}\cr
h(\phi)&=\frac{1}{N}\sum_{i=1}^N  \Big(\Re{(v_i)}\Re{( {\rm F}_i)}-\Im{(v_i)}\Im{( {\rm F}_i)}\Big)\qquad{\rm LH}
\label{hest}
\end{aligned}
\end{equation}
Because the phase $\phi$ is not known, the calculation of \Eq{hest} is carried out for a number of
values for $\phi$, typically 36.  Experimentally, a plot of $h$ as a function of $\phi$ is an extremely
smooth sine wave, even for data that is pure noise.  For that reason we take the maximum of such
a plot minus the minimum as our estimate $h_e$ of the amplitude $h_0$:
\begin{equation}
\begin{aligned}
h_e&=\frac{1}{2}\Big({\rm max}\big(h(\phi)\big)-{\rm min}\big(h(\phi)\big)\Big)
\label{h0}
\end{aligned}
\end{equation}


\subsection{Search Data Preparation}
\begin{figure}[!ht]
\begin{center}
\includegraphics[height=10cm]{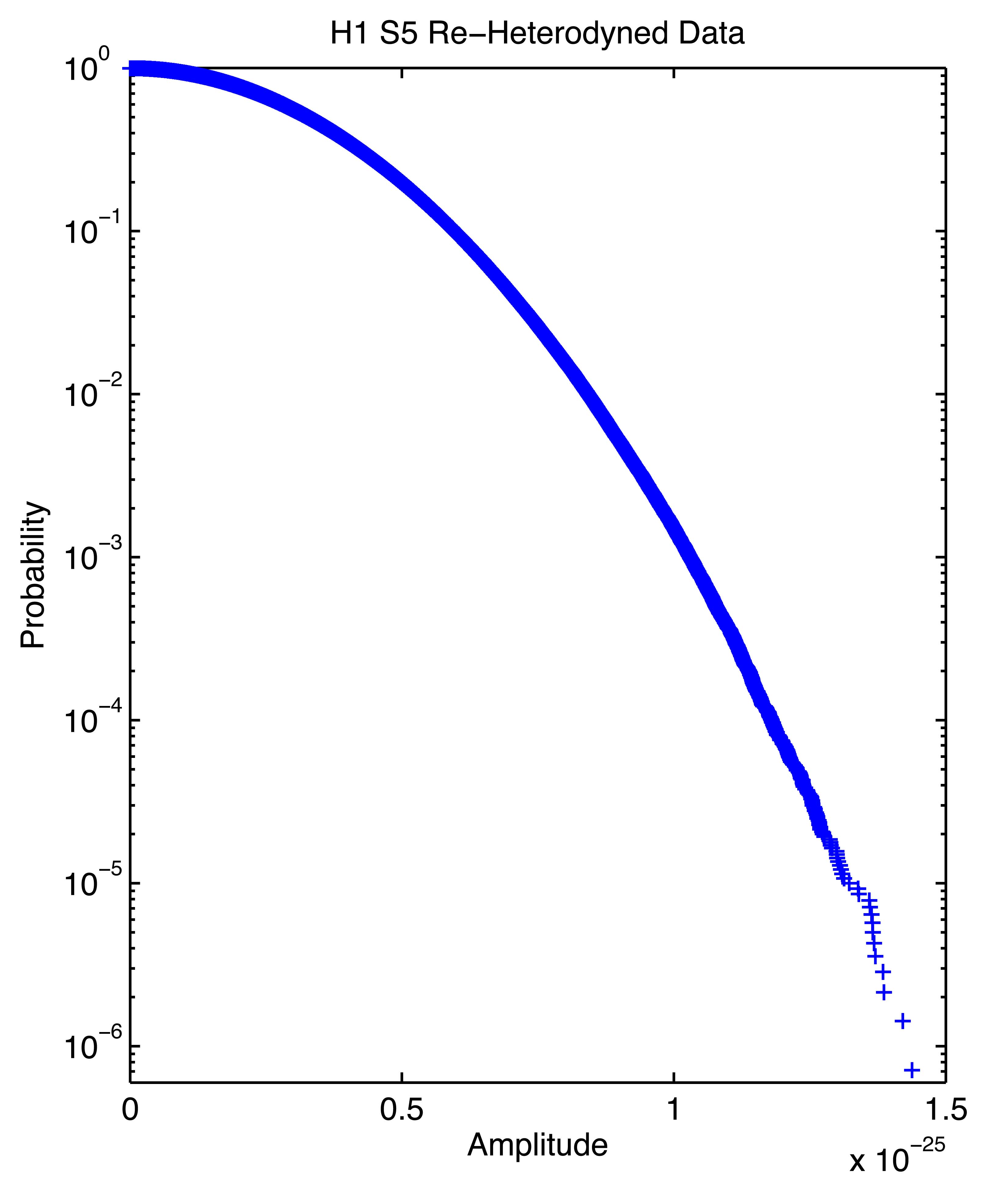} 
\caption{Cumulative distribution of detected amplitudes for 1,400,000 searches of H1 S5 data at frequencies spaced evenly
above and below the Crab ``beep'' frequency.  The form of this plot was chosen to emphasize the tail of the distribution.
\label{CumScan1400}}
\end{center}
\end{figure}

The process began with heterodyned data prepared with exquisite care by Matt Pitkin.  The largest outliers were
trimmed as described in Section \ref{PLC}.  At this point the standard deviation
of the complex data for H1 was $1.59\times 10^{-23}$.  Before attempting a search for any actual signal,
it was important to establish the background statistical properties of the data as it is reflected in accidental
correlations between the noise and the templates.  For this purpose we require a version of the received signal
in which there is no Crab signal.  \\This Catch-22 situation is solved in the following way:  The complex data are
re-heterodyned by multiplying each complex data sample by $e^{i\delta t}$ where $t$ is the sample time and 
$\delta$ is a frequency well within the passband of the heterodyned data.   This operation does not change 
the amplitude, but approximately orthogonalizes the signal with respect to the template.  
A data set of this type was prepared using frequencies from .008 Hz below the Crab
frequency to .008 Hz above, spaced by $1.16\times 10^{-8}$ Hz, giving a total of $1.4\times 10^6$ frequencies.  
Each frequency was re-heterodyned to base band and matched to both right-hand and left-hand rotation G4v templates.
The results are shown in \Fig{CumScan1400}.

\subsection{Signal Insertions}
\begin{figure}[!h]
\begin{center}
\begin{tabular}{c}
\includegraphics[height=7cm]{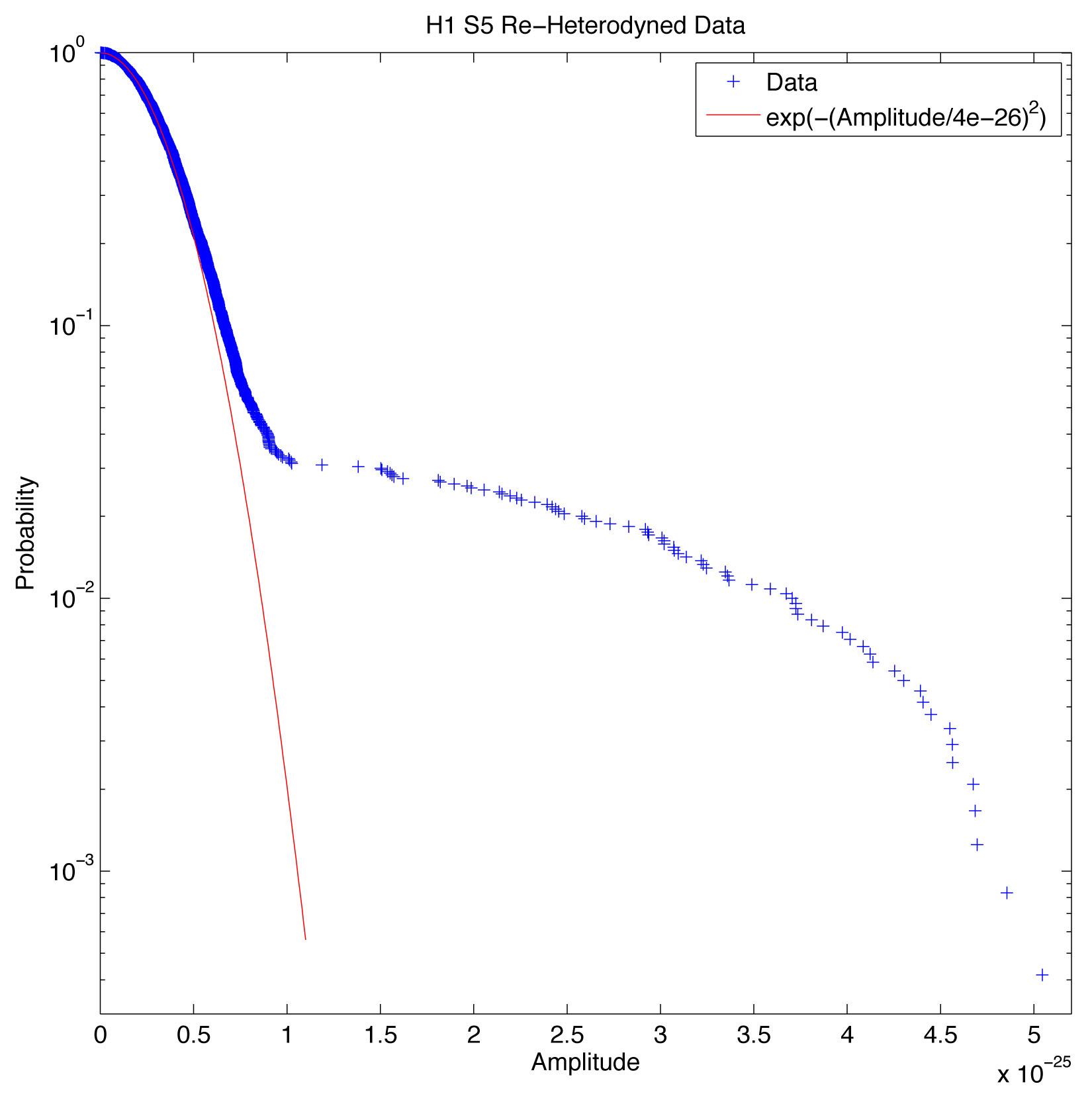}\hskip1in 
\includegraphics[height=7cm]{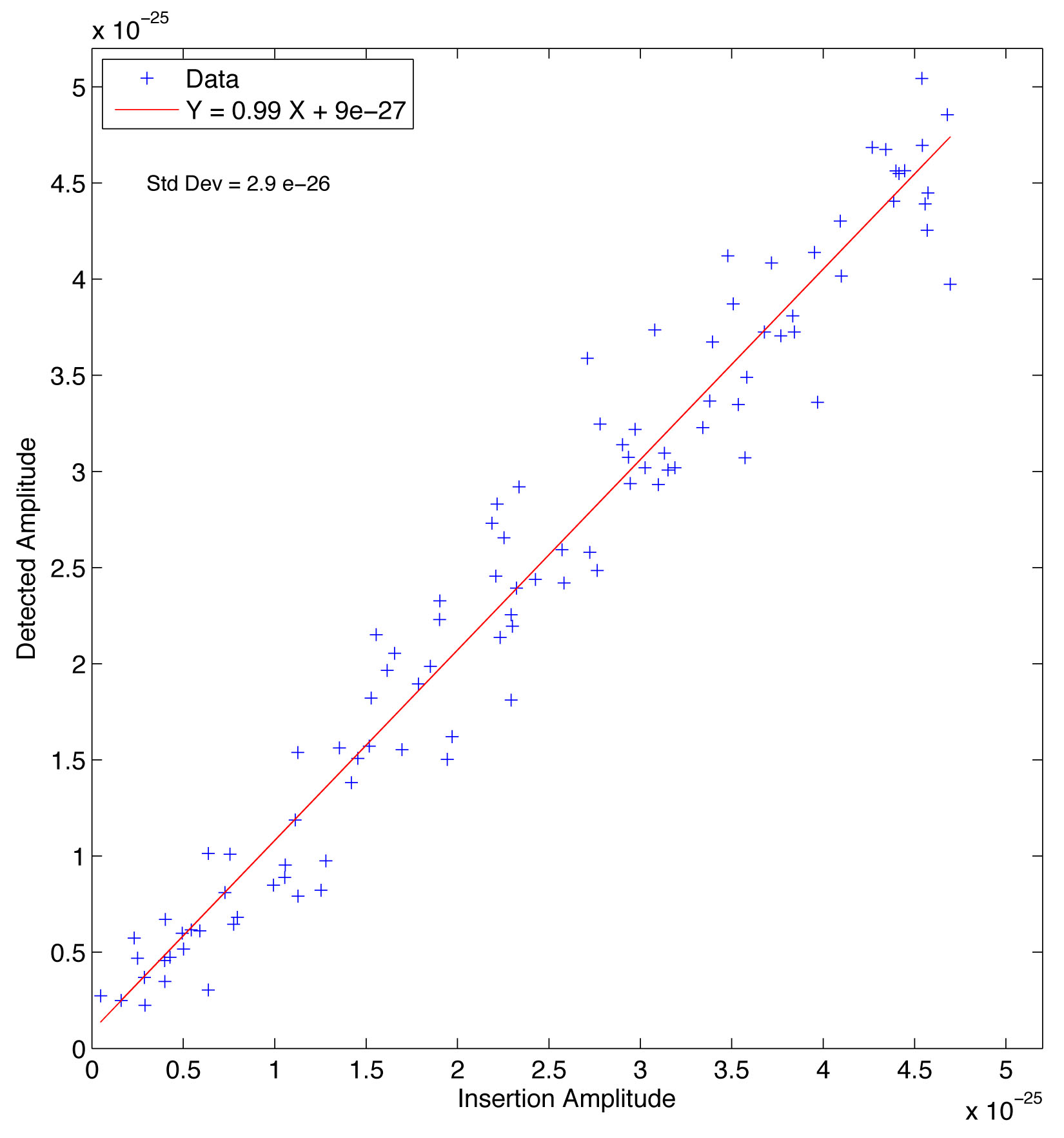} 
\end{tabular}
\caption{Left: Cumulative distribution of detected amplitudes for 2400 searches. 
2300 of these data sets are a subset of those shown in \Fig{CumScan1400} and contained no added signal.
100 additional sets contained synthesized signals randomly distributed in amplitude and phase.  Right: Comparison
of the amplitude recovered in the search {\it vs} the amplitude actually inserted
for the 100 additional data sets
\label{HCum_Random_Signals}}
\end{center}
\end{figure}

A set of 100 G4v Crab signals was prepared with amplitudes randomly distributed from zero to
3\% of the standard divination of the data, and phases randomly distributed from 0 to $2\pi$.
A subset of the original $1.4\times10^6$  data sets were augmented with 100 additional sets, to each of which was added
a different one of the synthesized  signals.  The results are shown in the left plot in \Fig{HCum_Random_Signals}. 
By keeping track of the amplitude of each insertion, we can compare it with the amplitude actually recovered
by the search of the data set containing that insertion, as shown in the right plot in \Fig{HCum_Random_Signals}.

These two plots give us two methods to estimate the probability that a detected amplitude
could have occurred by chance:  For example, the often-used rule-of-thumb that any point greater
than 5.326724 $\sigma$ has only one chance in ten million of arising by chance would place the
detection level at $1.635\times 10^{-25}$ using the right plot.  The same criterion
would place the level of detection at $1.615\times 10^{-25}$ using the expression for the red line in the left plot.
These detection thresholds are just above 1\% of the standard deviation of the original input signal.  

A much more comprehensive search process, including searches sensitive to both GR and G4v signals and method for
distinguishing between them is given in Isi {\it et.al.}\cite{Isi14}.  
There it is shown that ``For some combinations of detectors, sources, and signal strengths, G4v signals are invisible
to GR templates and vice versa.''   Thus the imminent detection of gravitational waves by Advanced LIGO and other
highly sensitive observatories will give us a powerful method for comparing theories of gravity in this new dynamic domain.


\section{Power Line Contamination}
\label{PLC}
\begin{figure}[!h]
\begin{center}
\begin{tabular}{c}
\includegraphics[width=9cm]{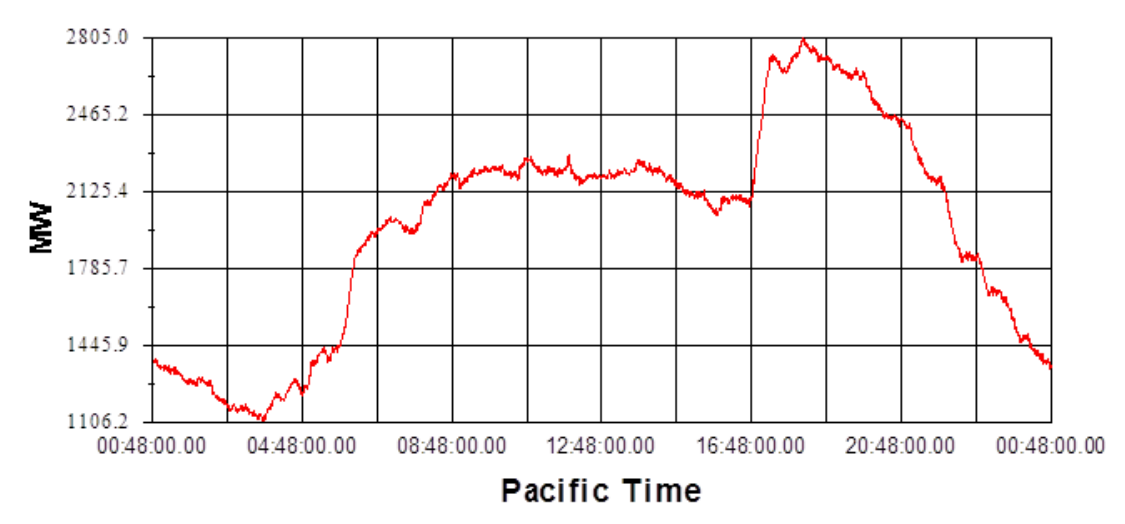} 
\includegraphics[width=9cm]{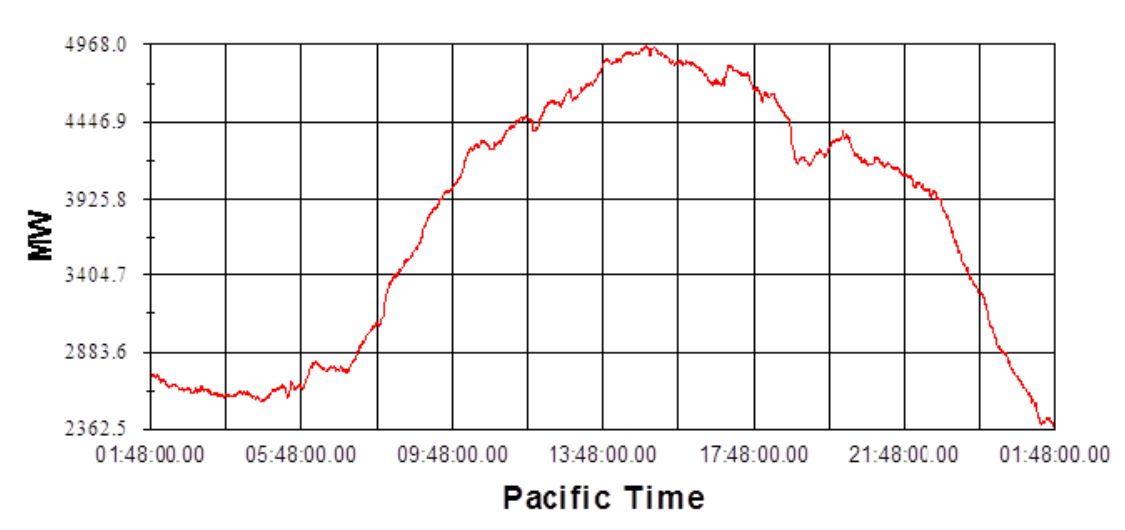} 
\end{tabular}
\caption{Load as a function of time for a region in the southwestern Unted States. 
The left curve is for one week-day in January and the right curve is for one week-day in August .
In both plots the daily maximum is more than twice the daily minimum. 
Notice the remarkable difference between the total power usage and the daily pattern in the two plots. 
This huge daily variation and seasonality of power usage is typical of many areas in the United States.
\label{LoadJanAug}}
\end{center}
\end{figure}

The LIGO Crab paper\cite{LIGOCrab08}  contains the following assertion:
``Although $2\nu$ [59.56 Hz] is close to the 60 Hz power line frequency,
it is sufficiently far away that the searches are relatively unaffected
by nonstationary components of the power line noise.''
It is the purpose of this section to examine the issue of power-line contamination in more detail.
The issue is complex for a number of reasons.  Although the long-term average frequency of the U.S. power grid
is held at 60 Hz to high accuracy, instantaneous deviations can occur in local areas due to sudden increases or decreases in load.
Power-line noise therefore has a strong daily variation due to the life and work patterns of our human society.  These time dependencies
are evident in the load curves kept by electric power providers, 
as illustrated by \Fig{LoadJanAug}.

Factories are started up every morning.  Lights come on every evening.   
 Induction motors run at a rotation rate lower than the synchronous speed
(60 or 30 Hz) by an amount dependent  on their load.  Under no-load conditions a high-quality induction motor
can run arbitrarily close to synchronous rate, progressively slowing as load is increased.  Florescent lights generate
large amounts of noise near 60 and 120 Hz, with considerable bandwidth due to the inconsistent firing point of
the plasma as the applied AC voltage waveform changes sign.  All of these phenomena make predicting the noise
properties near line frequency problematic.

\subsection{Statistical Properties of the S5 Crab Data}
The fifth LIGO science run (S5), started on 2005 November 4 and
ended on 2007 October 1. 
During this period the detectors (the 4 km and 2 km detectors at LIGO
Hanford Observatory, H1 and H2, and the 4 km detector at
the LIGO Livingston Observatory, L1) were at their design
sensitivities and had duty factors of 78\% for H1, 79\% for H2, and 66\% for L1.
No gravitational-wave signal was detected from the Crab Pulsar at a level above 1\% of the standard
deviation of the norm of the heterodyned data\cite{LIGOCrab08}.  Thus the outliers must be from causes other than the Crab,
and any correlation with the Crab gravitational-wave signal must be purely accidental.
For that reason we only consider the two most sensitive detectors H1 and L1.
\begin{figure}[!h]
\begin{center}
\begin{tabular}{c}
\includegraphics[height=8cm]{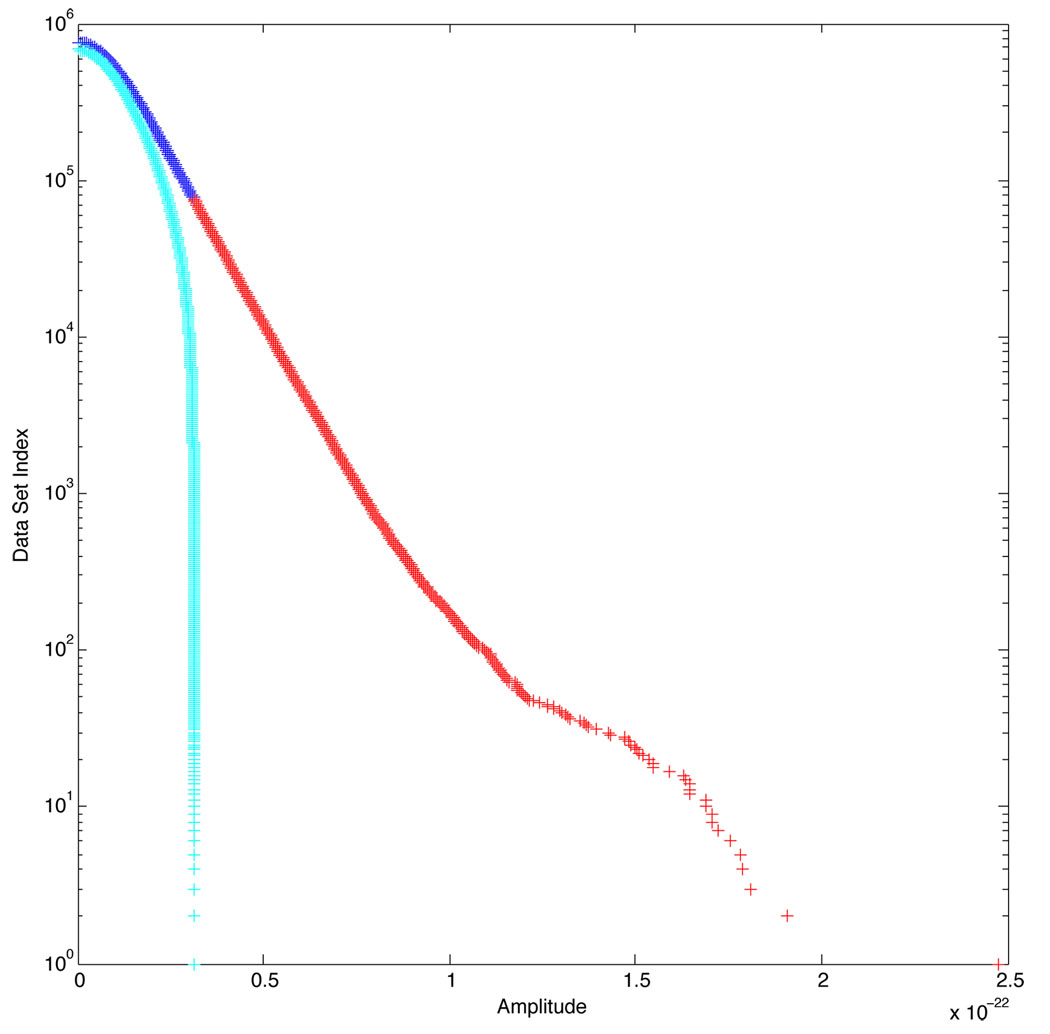}\hskip1in 
\includegraphics[height=8cm]{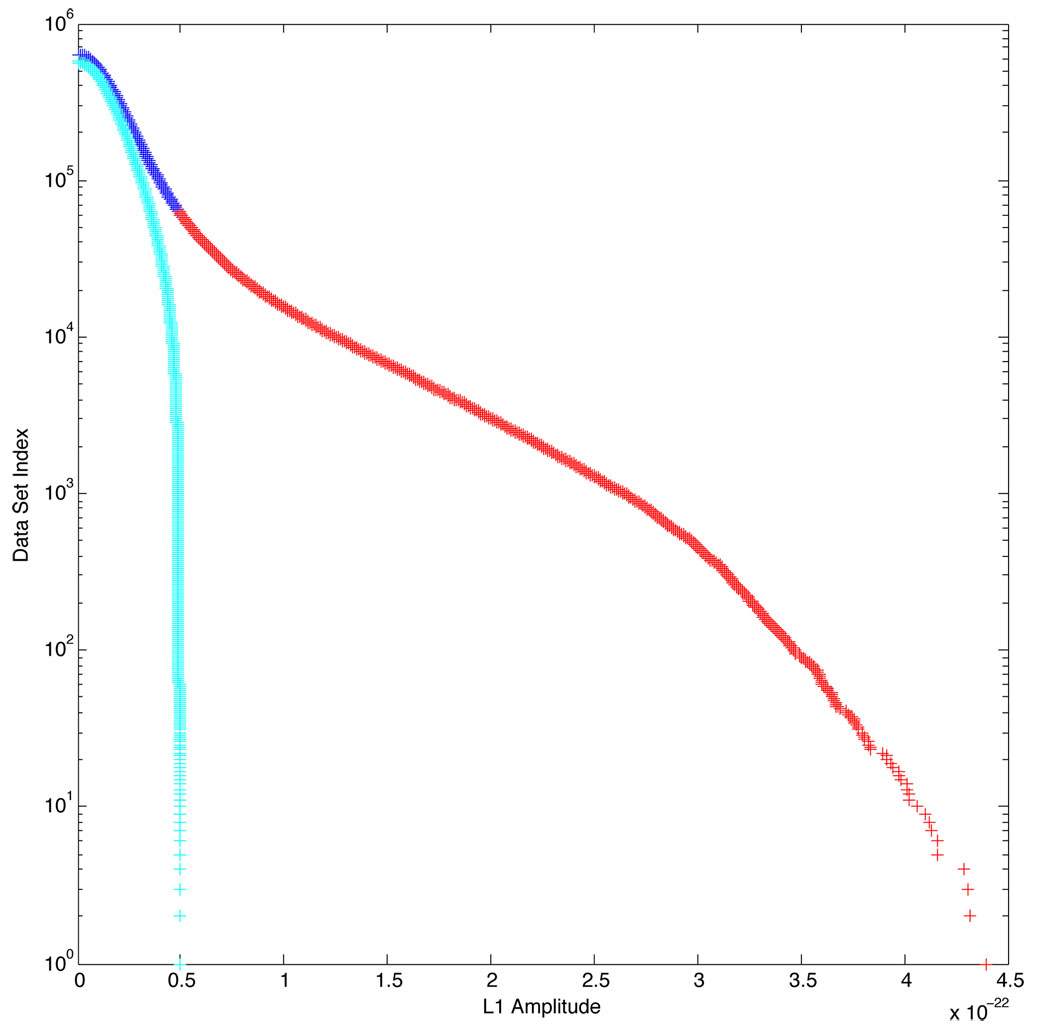} 
\end{tabular}
\caption{Cumulatve distribution functions of H1 (left) and L1 (right) heterodyned data norms from the S5 Crab search. 
In both plots the red and dark blue points are the full data set and the light blue points are for the same data with the
largest outliers (red points) removed.  The reduced sets contain 90\% of the heterodyned data,
corresponding to a duty factor of 70.2\% for H1 and 59.4\% for L1. 
\label{H1L1CDF}}
\end{center}
\end{figure}

The cumulative distribution functions of the heterodyned signal norm from these two detectors are shown in the \Fig{H1L1CDF}.
This plot format was chosen to best portray the behavior of the ``long tail'' of outliers.  
It is clear that the distribution is unusual, with a lot of weight carried by the outliers,
which could not have been due to the Crab signal.  Alan Weinstein suggested excluding the worst offenders\footnote{
``Trimming'' is the term often used in the statistical literature for this procedure.}
and re-analyzing the data to see if, perhaps, the outliers, normally associated with transient events, 
were also affecting the long-term spectral properties of the search. 

\subsection{Spectral Properties of the S5 Crab Data}
Any gravitational-wave signal from the Crab Pulsar will be manifest as a periodic variation of the received signal
with period 1 per sidereal day, with a strong component at 2 per sidereal day.  
Unfortunately, that period is within 1 part in 365 of 1 per solar day.

If power-line contamination were strictly periodic with solar-day period, we could separate it
cleanly from the sidereal-day period gravitational-wave signal by using one year's worth of data.
However we are not that fortunate for several reasons.
The S5 run was not quite a year in duration  
(201 days of data for H1, and 158 days of data for  L1).
Power-line load patterns vary a great deal on a yearly timescale.  
Summer entails a large air-conditioning load that is not present in the winter.
Daylight savings time shifts the phase of some human activity and not that of others, etc., etc.
These yearly variations, evident in \Fig{LoadJanAug},
modulate contamination signals, and thus create sidebands
shifted from the solar day period by exactly the same spacing as the sidereal day signal

{Because the Earth rotates around its axis in the same sense that it rotates around the Sun, the number of  solar days
in a year is one less than the number of sidereal days in a year.  Because of the trigonometric identity
$2\cos{A}\cos{B}=\cos{(A+B)}+\cos{(A-B)}$, any
signal periodic at one per solar day and amplitude modulated at one per year will have sidebands at
$\pm 1/({\rm days\ per\ year})$ from the one per solar day fundamental.  
The upper of these sidebands will therefore coincide with the sidereal day frequency.}

The spectral estimation process for data of this kind is problematic because the data
are not uniformly sampled. The nominal sampling rate after heterodyning
and low-pass filtering is 1 per minute. However there are many gaps in the
data where no samples exist. Standard Fourier transform techniques 
cannot be applied directly under these conditions. We were fortunate to contact
Dick Lyon of Google just after he had updated the Wikipedia page on Least
Squares Spectral Analysis. He suggested using the Lomb-Scargle algorithm,
which is specifically optimized for nonuniformly sampled data.
Matlab code and source reference are given in Section \ref{LSA}.

We begin by analyzing the Lomb-Scargle spectrum of each observatory in the neighborhood of the first
two harmonics of 1 per day, as shown in the following plots.  The most obvious difference
between the two observatories is that all the L1 harmonics are much larger than those from H1. 
The second thing to note is that all the spectra are changed in fundamental ways when
10\% of the data, made up of the largest outliers, have been removed.

\begin{figure}[!h]
\begin{center}
\begin{tabular}{c}
\includegraphics[height=7cm]{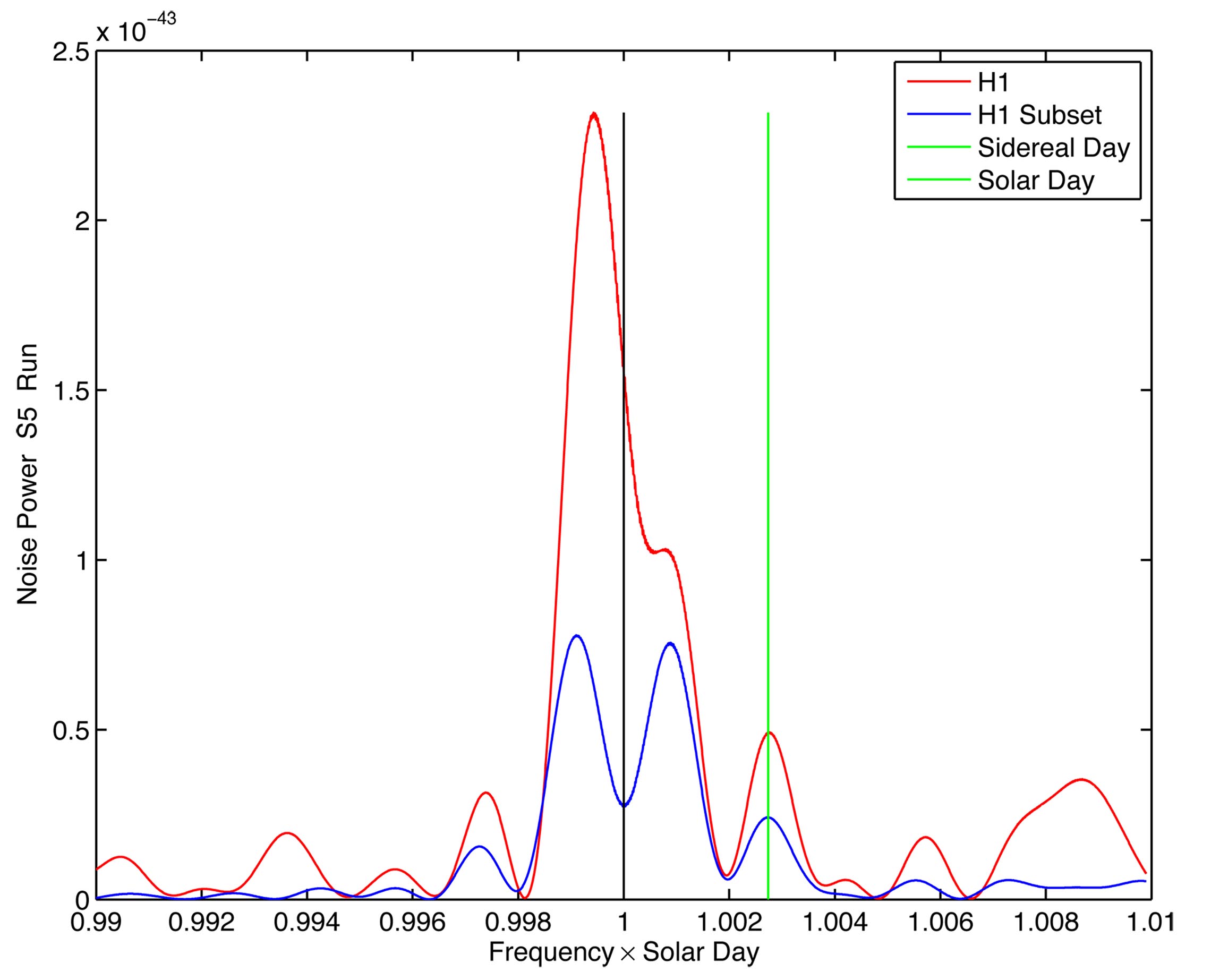}\hskip0.4in 
\includegraphics[height=7cm]{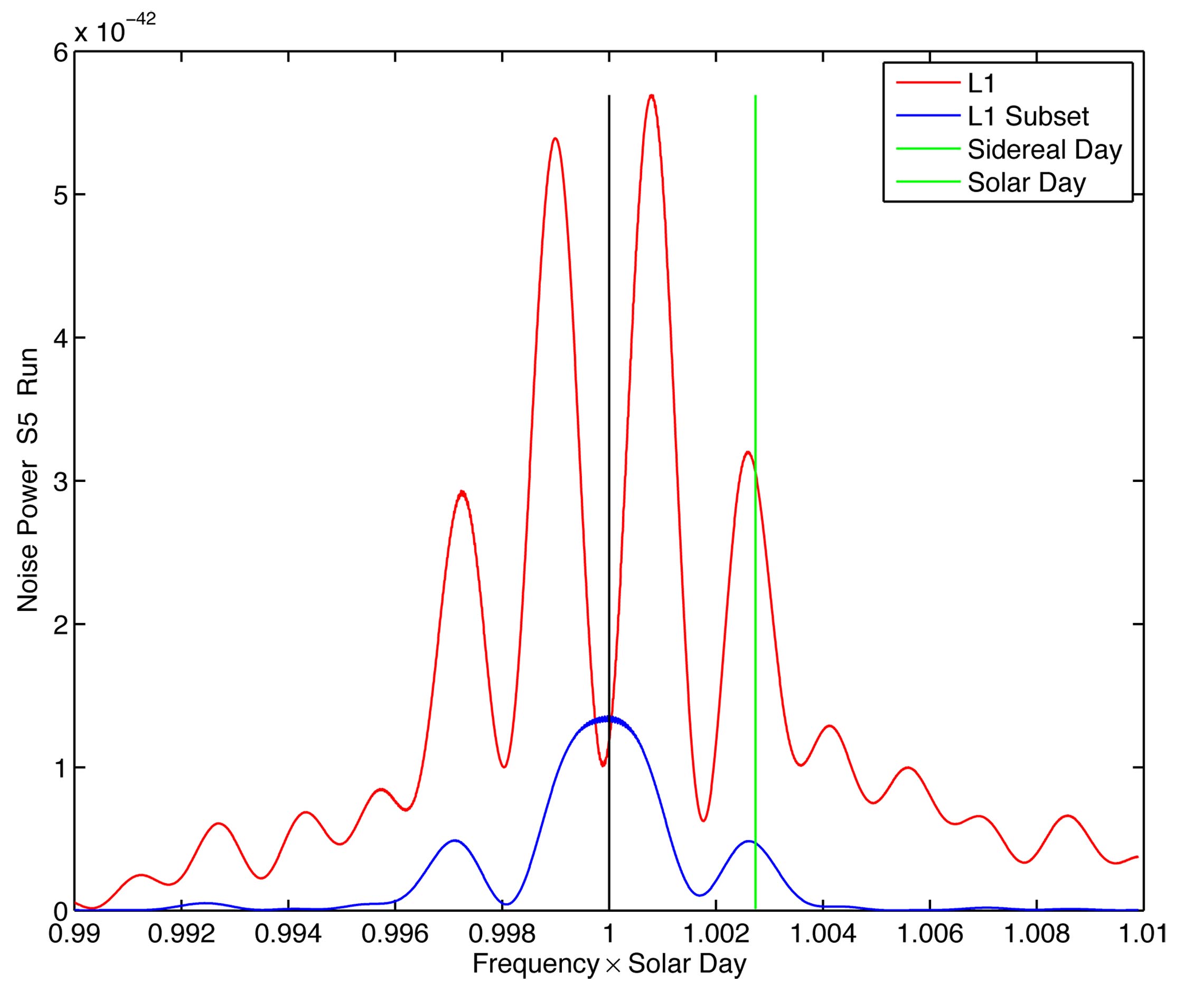} 
\end{tabular}
\caption{Spectrum of the norm of H1 (Left) and L1 (Right) heterodyned data in the neighborhood of one per day. 
The red curves correspond to the full datasets and the blue curve is from
the 90\% subsets shown in light blue in \Fig{H1L1CDF}.  The main peak is reduced considerably
when 10\% of the data, made up of the largest outliers, are removed. 
The peak around one per sidereal day, critical for any significant detection, is reduced by a factor of 2 in the H1 data
and by nearly a factor of 10 in the L1 data by this minor trimming process.
\label{H1L1Day}}
\end{center}
\end{figure}
\begin{figure}[!h]
\begin{center}
\begin{tabular}{c}
\includegraphics[height=7cm]{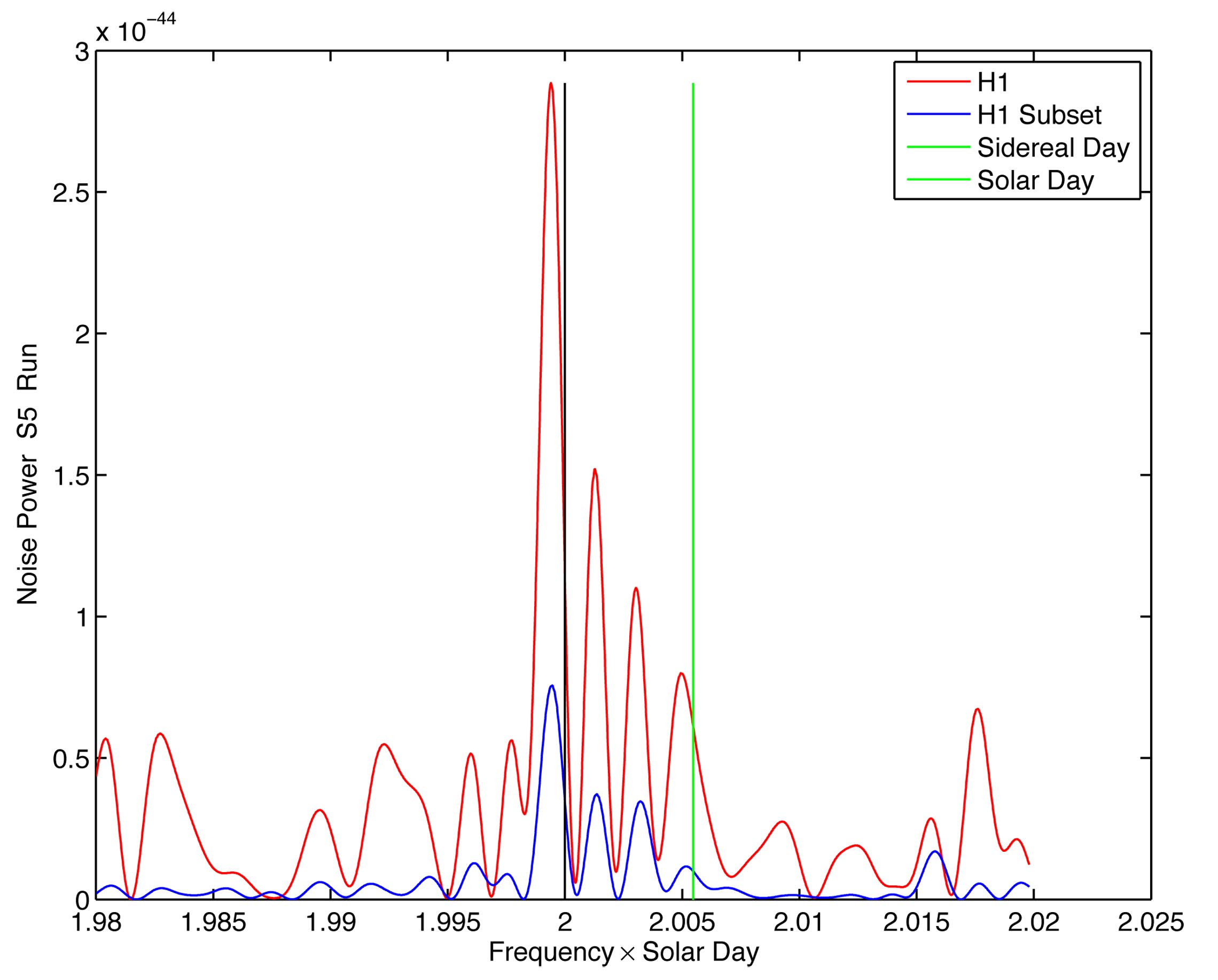}\hskip0.4in 
\includegraphics[height=7cm]{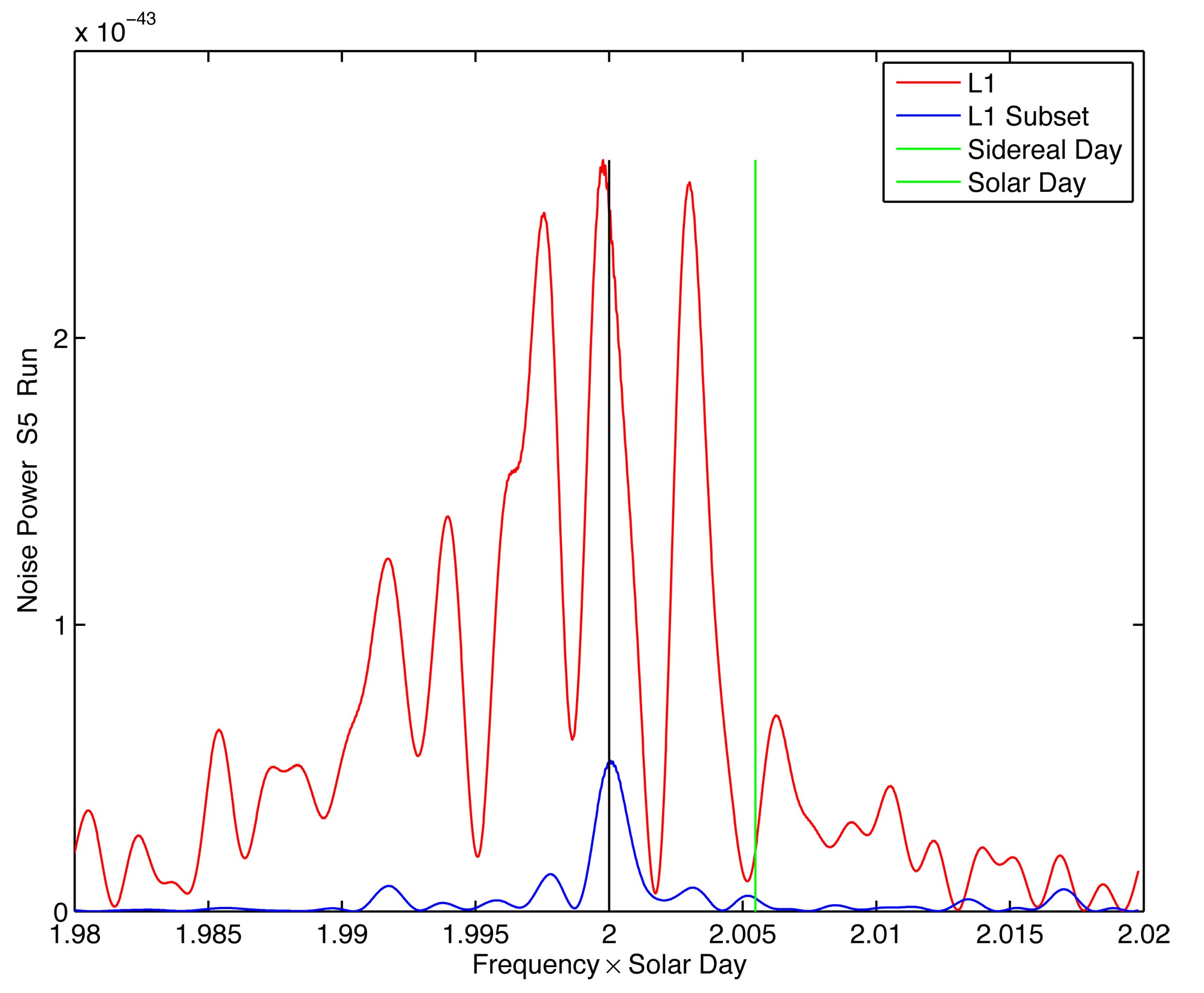} 
\end{tabular}
\caption{Spectrum of the norm of H1 (Left) and L1 (Right) heterodyned data in the neighborhood of two per day. 
The red curves correspond to the full datasets and the blue curves are from
the 90\% subsets shown in light blue in \Fig{H1L1CDF}.  The main peak is reduced considerably
when 10\% of the data, made up of the largest outliers, are removed. 
The peak around one per sidereal day, critical for any significant detection, is reduced by a factor of 2 in the H1 data
and by nearly a factor of 10 in the L1 data by this minor trimming process.
 The sidebands out to the sidereal second harmonic are greatly
reduced in the data sets that exclude the largest outliers. 
Note that the L1 spectrum in this region is about an order of magnitude larger than that of H1.
\label{H1L1HalfDay}}
\end{center}
\end{figure}

\vfil\eject
\subsection{Discussion}
There are many phenomena that have a daily period modulated on a yearly basis, 
the power line load being only one of them.
However the narrow (1/60 Hz) bandwidth of the Crab heterodyned data near 60 Hz makes power-related
noise particularly suspect.  
If this supposition is taken as a working hypothesis, it might be possible to find correlation between some
of the LIGO environmental monitors and the outliers identified in the present study.  On the other hand,
it is entirely possible that the Crab heterodyne is itself the most sensitive indicator of disturbances in
its narrow band of frequencies.  
There are an uncountable number of ways the power line can couple
into any experiment.  The LIGO interferometer, being the most sensitive instrument ever devised
by the human race, is vastly more prone to contamination than any other.  The consequence of this
situation is that no environmental monitor is likely to be anywhere near as sensitive to this contamination
as the interferometer itself.

\subsection{Hanford {\it vs} Livingston}
The integrity of the power grid near Hanford is particularly high.
The facility is near the center of one of the most extensive hydroelectric power developments in the world\cite{ColumbiaHydro}.
Not only do the nearby power plants have a large excess generating capacity, but, more important
in the present context, they provide a massive ``spinning reserve"---the term for synchronous, electrically coupled
angular momentum that lends stability to the phase of the power grid voltage in the face of load transients.
Perhaps nowhere else in the world is the power grid environment so favorable for a near-60 Hz search.
These circumstances lead us to interpret the remarkable superiority of the Hanford Crab data as being,
most probably, the result of its superior power-line integrity. 

No gravitational-wave signal was detected from the Crab Pulsar at a level above 1\% of the standard
deviation of the norm of the heterodyned data\cite{LIGOCrab08}.  Thus the outliers must be from sources other than the Crab,
and any correlation with the Crab gravitational-wave signal must be purely accidental.
This being the case, the transient outliers in the raw heterodyned data
can be discarded as described here with negligible impact on the statistical significance of the result due
to ``snooping the data''\cite{Mostafa12}.  Calculating the exact revision to the p-values
determined in the search due to this data cut is left to the experts.

We conclude that discarding at least 10\% of the data points, corresponding to the largest outliers 
in the heterodyned Crab data, is a logical and reasonable step in the Crab Pulsar search process, 
and that any effect on the integrity of the result due to this step is likely to be be negligible.

\vfil\eject
\section{Appendix - Transverse Theorem}
\label{AppendixTT}
We present this theorem for electromagnetic radiation.  
\\ In G4v, the relations for weak gravitational radiation are identical with two changes:
\begin{enumerate}
\item{The charge-current-density 4-vector is replaced by the energy-momentum 4-vector.}
\item{The coupling constant is $- G/c^2$ rather than ${1}/{4\pi\epsilon_0}$.}
\end{enumerate}
The Green's Function integral forms for the retarded scalar and vector potentials are:
\begin{equation}
\begin{aligned}
{\cal V}(t)&=\frac{1}{4\pi\epsilon_0}\int\frac{\rho(t-r/c)}{r}\, d{\rm vol}\qquad\qquad
{\vec A}(t)&=\frac{\mu_0}{4\pi}\int\frac{\vec J(t-r/c)}{r}\, d{\rm vol}
\label{GreenFunc}
\end{aligned}
\end{equation}
where $r$ is the distance from the source point to the point at which $\cal V$ and $\vec A$ are measured.

We consider a compact charge distribution that is changing with time, and inquire about
the general properties of the far-field radiation from that charge distribution.  By far-field
we mean that the distance $r$ from any point in the source to the point of observation is larger than
any dimension of the source by a large enough factor that, for any two points in the source,
$1/r$ is constant to within the accuracy required.  We choose our coordinate system so that the vector from the source to
the point of observation is in the $x$ direction.  We divide the source into thin slices
with planes of constant $x$, such that each slice of thickness $\delta x$ contains charge $q(x,t)$. 
We choose the origin $x=0$ such that its distance to the point of measurement is $R$. 
From \Eq{GreenFunc}, the potential ${\cal V}$ contributed by the charge $q(x,t)$ in a single slice
located at $x$ at time $t$ will be delayed by the propagation time $(R-x)/c$
\begin{equation}
\begin{aligned}
{\cal V}(t) =\frac{1}{4\pi\epsilon_0 R}q\big(t-(R-x)/c\big)=\frac{1}{4\pi\epsilon_0 R}q(t-R/c+x/c)
\label{Vslice}
\end{aligned}
\end{equation}
Therefore the potential ${\cal V}_1$ contributed by the charge $q_1(x,t)$ in a single slice located at $x$ is
\begin{equation}
\begin{aligned}
{\cal V}_1(t') =\frac{1}{4\pi\epsilon_0 R}\,q_1(t)
\label{Vslice1}
\end{aligned}
\end{equation}
where $t=t'-R/c+x/c$.
The potential at the same observation point from the immediately adjacent slice (2) located at
$x+\delta x$ is
\begin{equation}
\begin{aligned}
{\cal V}_2(t') =\frac{1}{4\pi\epsilon_0 R}\,q_2\big(t+\delta x/c\big)
\label{Vslice2}
\end{aligned}
\end{equation}
So the total potential $V$ at the point $R$ will be

\begin{equation}
\begin{aligned}
{\cal V}&={\cal V}_1+{\cal V}_2 
=\frac{1}{4\pi\epsilon_0 R}\Big(q_1(t)+q_2(t+\delta x/c)\Big)\cr
&=\frac{1}{4\pi\epsilon_0 R}
\left(q_1(t)+q_2(t)+\frac{\delta x}{c}\,\frac{\partial{q_2}}{\partial t}\right)\cr
\frac{\partial{\cal V}}{\partial t'}
&=\frac{1}{4\pi\epsilon_0 R}\Big(\frac{\partial{q_1}}{\partial t}
+\frac{\partial{q_2}}{\partial t}+\frac{\delta x}{c}\,\frac{\partial^2{q_2}}{\partial t^2}\Big)
\label{dRslice}
\end{aligned}
\end{equation}
If we consider these two adjacent slices in isolation, the charge on the first will be decreased 
and that on the second will be increased by a current $I$ crossing
the boundary between them in the $+x$ direction.  

By conservation of charge
\begin{equation}
\begin{aligned}
\frac{\partial{q_2}}{\partial t} &=-\frac{\partial{q_1}}{\partial t} =I
\end{aligned}
\label{dqdtslice}
\end{equation}
which, with \Eq{dRslice} gives
\begin{equation}
\begin{aligned}
\frac{\partial{\cal V}}{\partial t'}
&=\frac{\delta x}{4\pi\epsilon_0 R c}\,\frac{\partial^2{q_2}}{\partial t^2}
\label{dVdtslice}
\end{aligned}
\end{equation}
The potential $\cal V$ is propagating in the $+x$ direction at velocity $c$.  
Its functional form is therefore ${\cal V}=f(t-R/c)$.  The derivatives are thus related by
\begin{equation}
\begin{aligned}
\frac{\partial{\cal V}}{\partial R}&=
-\frac{1}{c}\frac{\partial{\cal V}}{\partial t'}
=-\frac{\delta x}{4\pi\epsilon_0 c^2 R}\,\frac{\partial^2{q_2}}{\partial t^2}
=-\frac{\mu_0\,\delta x}{4\pi R}\,\frac{\partial^2{q_2}}{\partial t^2}
\label{dVdRslice}
\end{aligned}
\end{equation}
where we have used the fact that $\mu_0\epsilon_0=1/c^2$.\\ 
By \Eq{GreenFunc} and  \Eq{dqdtslice}, the vector potential at point $R$ will be
\begin{equation}
\begin{aligned}
A(R,t')&=\frac{\mu_0}{4\pi R}I(t)\delta x=\frac{\mu_0\delta x}{4\pi R}\,\frac{\partial{q_2}}{\partial t}\cr
\frac{\partial A}{\partial t}&=\frac{\mu_0\delta x}{4\pi R}\,\frac{\partial^2{q_2}}{\partial t^2}
\label{Aslice}
\end{aligned}
\end{equation}
From \Eq{dVdRslice} and From \Eq{Aslice}, we conclude that the total electric field $\vec {\cal E}$ in the direction of propagation vanishes:
\begin{equation}
\begin{aligned}
\frac{\partial A}{\partial t}&=-\frac{\partial{\cal V}}{\partial R}\qquad\Rightarrow\qquad
\vec {\cal E} =-\nabla{\cal V} -\frac{\partial\vec A}{\partial t}=0
\label{Et0}
\end{aligned}
\end{equation}
We can make up any compact charge distribution out of a superposition of pairs of slices
like we have just analyzed.  Because the Green's Functions (\Eq{GreenFunc}) are linear, if the longitudinal
electric field is zero for each pair of slices individually, it will be zero for the sum as well.

We have thus arrived at a very important result for an electromagnetic wave propagating in free space
far from its source:

{\bf Transverse Theorem}: The electromagnetic wave propagating in free space due to 
a compact source has the following property:

\noindent
{\it The time derivative of the longitudinal vector potential just cancels the spatial
gradient of the scalar potential; therfore the electric field in the direction of propagation Zero.}

The far-field radiation from any source approaches a plane wave arbitrarily closely at large distances from
its source.  The scalar potential $\cal V$ in a plane wave is constant within any plane normal to the
direction of propagation.  It follows from the transverse theorem that the only attribute of the wave
that can affect charges is the
transverse vector potential $A_\perp$, and the only non-zero component of the electric field
is ${\cal E}_\perp=-\partial A_\perp/\partial t$.  Plane electromagnetic waves are thus said to be
Transverse Waves.  From the Green's Function (\Eq{GreenFunc}) it follows that
we can find the components of the vector potential that interact with matter by considering only
the components of source current normal to the direction of propagation.  This deep result enables
an enormous simplification in calculating any far-field radiation pattern.  The special case of
this result for an electric dipole source is shown in CE 4.16.

This conclusion is with regard to propagating wave solutions.  It obviously does not apply to the static electric
field of a charge distribution with a net charge that is not varying with time.

\vfil\eject

\section{Appendix - Lomb-Scargle Algorithm}
\label{LSA}
Matlab code for the Lomb-Scargle algorithm as given in:

{\bf Ice Ages and Astronomical Causes:\\
 Data, spectral analysis and mechanisms}\\
 By Richard A. Muller, Gordon J. MacDonald\begin{verbatim}
http://books.google.com/books?id=P8ideTkMQisC&pg=PA289&dq=spectral
+lomb+scargle&hl=en#v=onepage&q=spectral%20lomb%20scargle&f=false
\end{verbatim}

\begin{verbatim}
function power=lomb(t,y,freq)
% Lomb-Scargle periodogram
    nfreq=length(freq);
    fmax=freq(nfreq);
    fmin=freq(1);
    power=zeros(nfreq,1);
    f4pi=freq*4.*pi;
    pi2=2*pi;
    n=length(y);
    cosarg=zeros(n,1);
    sinarg=zeros(n,1);
    argu=zeros(n,1);
    var=cov(y); %variance
% subtract mean
    yn=y-mean(y);
    for fi=1:nfreq;
        sinsum=sum(sin(f4pi(fi)*t));
        cossum=sum(cos(f4pi(fi)*t));
        tau=atan2(sinsum,cossum);
        argu=pi2*freq(fi)*(t-tau);
        cosarg=cos(argu);
        cosnorm=sum(cosarg.*cosarg);
        cfi=sum(yn.*cosarg);
        sinarg=sin(argu);
        sfi=sum(yn.*sinarg);
        sinnorm=sum(sinarg.*sinarg);
        power(fi)=(cfi*cfi/cosnorm+sfi*sfi/sinnorm);
        %For normalized output
        %power(fi)=(cfi*cfi/cosnorm+sfi*sfi/sinnorm)/(2*var); 
    end\end{verbatim}
   The plots in this document were made without normalizing the output\\
    in order to compare the noise magnitude at H1 and L1 observatories.
    
{\bf Acknowledgements:}  \\I am indebted to Yaser Abu-Mostafa, Patrick Ennis, Mike Godfrey, Peter Goldreich, Mike Gottlieb, 
Bart Huxtable, Max Isi, Dick Lyon, Sanjoy Mahajan, Nathan Mead, Alonso Rodriguez, Rahul Sarpeshkar,
Jamil Tahir-Kheli, Kip Thorne, Lloyd Watts, and Alan Weinstein 
for many discussions and suggestions.\\
The exquisite heterodyne work is due to Matt Pitkin and Grahm Woan.  
    \\None of these gentlemen should be blamed for remaining defects in this communication.
    
\vfil\eject

\end{document}